\documentstyle[11pt,emulateapj]{article}

\def\p{\partial}

\def\nab{\mbox{\boldmath $\nabla$}}

\def\rb{\bar{\rho}}
\def\tb{\bar{T}}
\def\sb{\bar{S}}

\def\vph{\hat{v}_{\phi}}

\def\vrr{\tilde{v}_r} 
\def\vtr{\tilde{v}_{\theta}}
\def\vphr{\tilde{v}_{\phi}}
\def\vvr{\tilde{v}}

\newcommand{\del}{\mbox{$\nabla$}}

\def\vec#1{\mbox{\boldmath$\displaystyle#1$}}
\newcommand{\dotp}{\vec{\times}}

\begin{document}

\submitted{submitted June 8 2001, accepted December 14 2001}

\title{TURBULENT CONVECTION UNDER THE INFLUENCE OF ROTATION: \\
      SUSTAINING A STRONG DIFFERENTIAL ROTATION}

\author{Allan Sacha BRUN and Juri TOOMRE}

\affil{JILA and Department of Astrophysical and Planetary Sciences,
University of Colorado, Boulder, CO 80309-0440, USA.\\}

\begin{center}
sabrun@solarz.colorado.edu, jtoomre@jila.colorado.edu
\end{center}

\begin{abstract}

The intense turbulence present in the solar convection zone is a major
challenge to both theory and simulation as one tries to understand the
origins of the striking differential rotation profile with radius and
latitude that has been revealed by helioseismology.  The differential
rotation must be an essential element in the operation of the solar
magnetic dynamo and its cycles of activity, yet there are many aspects
of the interplay between convection, rotation and magnetic fields that
are still unclear.  We have here carried out a series of 3--D numerical
simulations of turbulent convection within deep spherical shells using our
anelastic spherical harmonic (ASH) code on massively parallel
supercomputers.  These studies of the global dynamics of the solar
convection zone concentrate on how the differential rotation and
meridional circulation are established.  We have addressed two issues
raised by previous simulations with ASH.  Firstly, can solutions be
obtained which possess the apparent solar property that the angular
velocity $\Omega$ continues to decrease significantly with latitude as
the pole is approached?  Prior simulations had most of their rotational
slowing with latitude confined to the interval from the equator to
about 45$^{\circ}$.  Secondly, can a strong latitudinal angular
velocity contrast $\Delta \Omega$ be sustained as the convection
becomes increasingly more complex and turbulent?  There was a tendency
for $\Delta \Omega$ to diminish in some of the turbulent solutions that
also required the emerging energy flux to be invariant with latitude.

In responding to these questions, five cases of increasingly turbulent
convection coupled with rotation have been studied along two paths in
parameter space.  We have achieved in one case the slow pole behavior
comparable to that deduced from helioseismology, and have retained in
our more turbulent simulations a consistently strong $\Delta \Omega$.
We have analyzed the transport of angular momentum in establishing such
differential rotation, and clarified the roles played by Reynolds
stresses and the meridional circulation in this process.  We have found
that the Reynolds stresses are crucial in transporting angular momentum
toward the equator.  The effects of baroclinicity (thermal wind) have
been found to have a modest role in the resulting mean zonal flows.
The simulations have produced differential rotation profiles within the
bulk of the convection zone that make reasonable contact with ones
inferred from helioseismic inversions, namely possessing a fast
equator, an angular velocity difference of about 30\% from equator to
pole, and some constancy along radial lines at mid-latitudes.  Future
studies must address the implications of the tachocline at the base of
the convection zone, and the near-surface shear layer, upon that
differential rotation.

\end{abstract}


\section{INTRODUCTION}

The solar turbulent convection zone has striking dynamical properties
that continue to challenge basic theory.  The most fundamental issues
involve the solar rotation profile with latitude and depth, and the
manner in which the 22-year cycles of solar magnetic activity are
achieved.  These two issues are closely interrelated, for the global
dynamo action is likely to be very sensitive to the angular velocity
$\Omega$ profiles realized by convection redistributing angular
momentum within the deep zone.  Both dynamical topics touch on the
seeming inconsistency that turbulence can be both highly intermittent
and chaotic on smaller spatial and temporal scales, yet exhibit
large-scale ordered behavior (cf.  Brummell, Cattaneo \& Toomre 1995).
The differential rotation profile established by the turbulent
convection, though strong in contrast, is remarkably smooth; the
global-scale magnetic activity is orderly, involving sunspot eruptions
with very well defined rules for field parity and emergence latitudes
as the cycle evolves.  The wide range of dynamical scales of turbulence
present in the solar convection zone yield severe challenges to both
theory and simulation:  the discernible structures range from granules
($\sim 10^3$ km or 1 Mm in horizontal size), to supergranules ($\sim$30
Mm), to possible patterns of giant cells comparable to the overall
depth of that zone ($\sim$200 Mm, or nearly 30\% by radius).  Given
that the dissipation scales are on the order of 0.1 km or smaller, the
solar turbulence encompasses at least six orders of magnitude for each
of the three physical dimensions.  The largest current 3--D turbulence
simulations can resolve about three orders of magnitude in each
dimension.  Yet despite the vast difference in the range of scales
dynamically active in the sun and those accessible to simulations, the
latter have begun to reveal basic self-ordering dynamical processes
yielding coherent structures that appear to play a crucial role in the
global differential rotation and magnetic dynamo activity realized in
the sun.

\begin{figure*}[!htp]
\centerline{\includegraphics[width=0.95\linewidth]{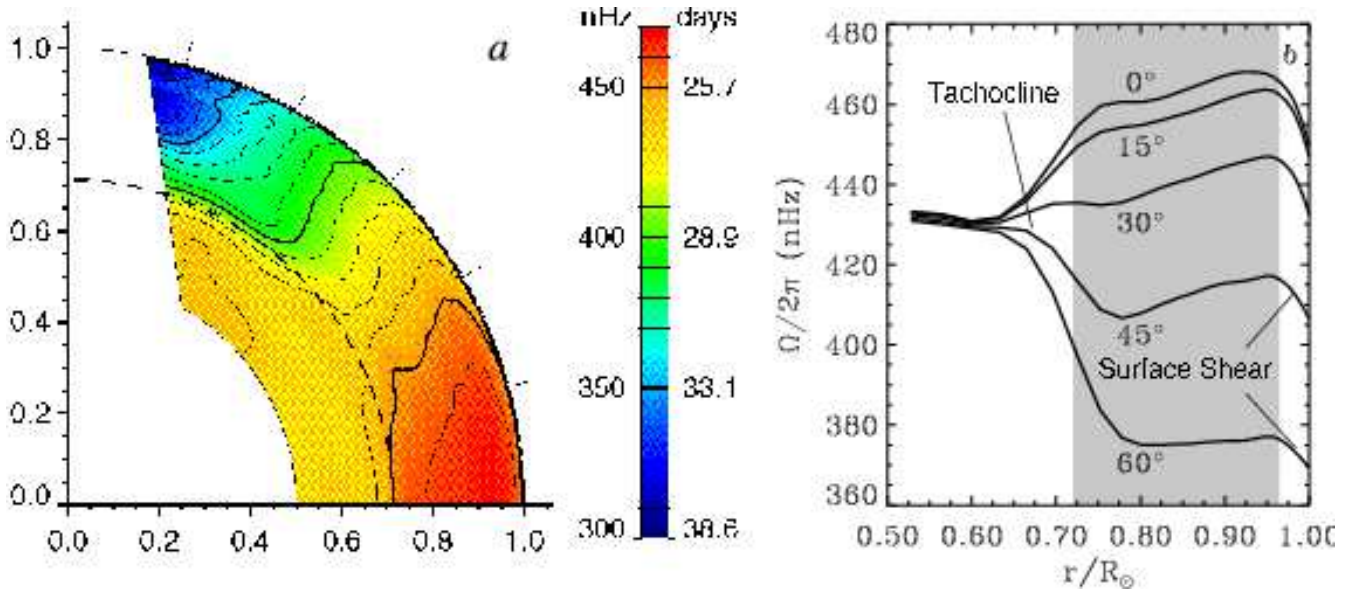}}

\caption[]{\label{fig1} ~($a$) Angular velocity profile
$\Omega/2\pi$ with radius and latitude as deduced from helioseismology
using
SOI--MDI data, with red tones indicating fast rotation and blue-green
the slowest rotation [adapted from Schou et al. 1998].  ~($b$) Time-averaged
rotation rates from five years of GONG helioseismic data, plotted
against radius at different latitudes. The surface shear layer and the
tachocline at the base of the convective zone are indicated, as well as 
the zone covered by our computational domain (grey area) [adapted from Howe et al. 2000b].}
\end{figure*}

It has long been known by tracking surface features that the surface of
the sun rotates differentially (e.g. Ward 1966, Sch\"ussler 1987):  there is
a smooth poleward decline in the angular velocity $\Omega$, the
rotation period being about 25 days in equatorial regions and about 33
days near the poles.  Helioseismology, which involves the study of the
acoustic $p$-mode oscillations of the solar interior (e.g. Gough \&
Toomre 1991), has provided a remarkable new window for studying
dynamical processes deep within the sun.  This has been enabled by
nearly continuous helioseismic observations provided from the SOHO
spacecraft with the high-resolution Michelson Doppler Imager (SOI--MDI)
(Scherrer et al. 1995) and from the ground-based Global Oscillation
Network Group (GONG) set of six related instruments (Harvey et al.
1996).  The helioseismic findings about differential rotation deeper
within the sun have turned out to be revolutionary, for they are unlike
any anticipated by convection theory prior to such probing of the
interior of a star.  Helioseismology has revealed that the rotation
profiles obtained by inversion of frequency splittings of the $p$ modes
(e.g. Libbrecht 1989, Thompson et al. 1996, Schou et al. 1998, Howe et
al. 2000b) have the striking behavior shown in Figure \ref{fig1}.  The
variation of angular velocity $\Omega$ observed near the surface, where
the rotation is considerably faster at the equator than near the poles,
extends through much of the convection zone with relatively little
radial dependence.  Thus at mid-latitudes $\Omega$ is nearly constant
on radial lines, in sharp contrast to early numerical simulations of
rotating convection in spherical shells (e.g. Gilman \& Miller 1981,
Glatzmaier 1987) that suggested that $\Omega$ should be nearly constant
on cylinders aligned with the rotation axis and decreasing inward
on the equatorial plane. Another striking feature is the region of strong 
shear at the base of the convection zone, now known as the tachocline, 
where $\Omega$ adjusts to apparent solid body rotation in the deeper radiative
interior.  Whereas the convection zone exhibits prominent differential
rotation, the deeper radiative interior does not; these two regions are
joined by the complex shear of the tachocline.  There is further a thin
shear boundary layer near the surface in which $\Omega$ increases with
depth at intermediate and high latitudes.

The tachocline has been one of the most surprising discoveries of
helioseismology, especially since its strong rotational shear affords a
promising site for the solar global dynamo.  Such a tachocline was not
anticipated, and current theoretical approaches to explain its presence
are still only innovative sketches (Spiegel \& Zahn 1992; Gough \&
McIntyre 1998; Charbonneau, Dikpati \& Gilman 1999).  Helioseismology
has also recently detected prominent variations in the rotation rate
near the base of the convective envelope, with a period of 1.3 years
evident at low latitudes (Howe et al. 2000a; Toomre et al. 2000).
These are the first indications of dynamical changes close to the
presumed site of the global dynamo as the cycle advances.  Such a
succession of developments from helioseismology provide both a
challenge and a stimulus to theoretical work on solar convection zone
dynamics.

Seeking to understand solar differential rotation and magnetism
requires 3--D simulations of convection in the correct full spherical
geometry.  However, the global nature of such solutions represent a
major computational problem, given that the largest scale is pinned and
only three orders of magnitude smaller in scale can be represented.
Much of the small-scale dynamics in the sun dealing with
supergranulation and granulation are by necessity then largely
omitted.  The alternative is to reduce the fixed maximum scale by
studying smaller localized domains within the full shell and utilizing
the three orders of magnitude to encompass the dynamical range of
turbulent scales.  There are clear tradeoffs: ~the global models
operate in the correct geometry yet struggle to encompass enough of a
dynamical range to admit fully turbulent solutions, whereas the local
models are able to study intensely turbulent convection but only within
a particular limited portion of the full domain.  Both approaches are
needed, and the efforts are complementary, as reviewed in detail by
Gilman (2000) and Miesch (2000).  Highly turbulent but localized 3--D
portions of a convecting spherical shells are being studied to assess
transport properties and topologies of dynamical structures (e.g.
Brandenburg et al. 1996; Brummell et al. 1996, 1998; Porter \& Woodward
2000; Robinson \& Chan 2001), of penetration into stable domains below
(Brummell et al. 2001, Porter \& Woodward 2001), of effects of
realistic near-surface physics on granulation and supergranulation
(e.g. Stein \& Nordlund 1998), and of dynamo processes and magnetic
transport by the convection (e.g. Cattaneo 1999, Tobias et al.  2001).
Without recourse to direct simulations, the angular momentum and energy
transport properties of turbulent convection have also been considered
using mean-field approaches to derive second-order correlations (the
Reynolds stresses and anisotropic heat transport) under the assumption
of separability of scales.   Although such procedures involve major
uncertainties, the resulting angular momentum transport, which is
described by mechanisms such as the so-called $\Lambda$ effect, have
served to reproduce the solar meridional circulation (e.g. Durney 1999,
2000) and differential rotation (e.g.  Kichatinov \& R\"{u}diger 1995).
Various other states can be achieved by adjusting parameters.

Initial studies of convection in full spherical shells to assess
effects of rotation with correct account of geometry (e.g. Gilman \&
Miller 1981; Glatzmaier \& Gilman 1982; Glatzmaier 1985, 1987; Sun \&
Schubert 1995) have set the stage for our efforts to study more
turbulent flows using new numerical codes designed for the massively
parallel computer architectures that are enabling such major
simulations.  We here report on our continuing studies with the
anelastic spherical harmonic (ASH) code (Clune et al.  1999) to examine
the $\Omega$ profiles established within the bulk of the solar
convection zone by turbulent convection, building on the progenitor
work by Miesch et al. (2000), Elliott, Miesch \& Toomre (2000), and
Brun \& Toomre (2001). We also recognize the recent modelling of 
convection in  spherical shells by Takehiro \& Hayashi (1999) and 
Grote \& Busse (2001).

The simulations reported in Miesch et al. (2000) and Elliott et al.
(2000) have revealed the richness and complexity of compressible
convection achieved in rotating spherical shells.  Most of the
resulting angular velocity profiles in the seven simulations considered
have begun to make substantial contact with the helioseismic
deductions within the bulk of the solar convection zone. These possess
 fast equatorial rotation (prograde), substantial
$\Omega$ contrasts with latitude, and reduced tendencies for rotation
to be constant on cylinders. The simulations with ASH
 have not yet sought to deal with questions of the
near-surface rotational shear layer nor with the formation of a
tachocline near the base of the convection zone.
 These studies have revealed that to achieve
fast equators it is essential that parameter ranges be considered in
which the convection senses strongly the effects of rotation, which
translates into having a convective Rossby number less than unity for
large Taylor numbers.  Such rotationally-constrained convection
exhibits downflowing plumes that are tilted away from the local radial
direction, resulting in velocity correlations and thus Reynolds
stresses that are found to have a significant role in the
redistribution of angular momentum.  This seems to provide paths to
realize solar-like $\Omega$ profiles.  Further, it is desirable to
impose thermal boundary conditions at the top of the domain that
enforce the constancy of emerging flux with latitude in order to be
consistent with what appears to be observed.

We wish to focus on two
outstanding issues raised by the prior simulations with ASH that need particular
attention concerning the differential rotation established within the
bulk of the solar convection zone.  As {\sl Issue 1}, the
helioseismic inferences in Figure 1 emphasize that $\Omega$ in the sun
appears to decrease significantly with latitude even at mid and high
latitudes, a property which has been difficult to attain in the prior
seven simulations.  The substantial latitudinal decrease in angular
velocity, say $\Delta \Omega$, in the models is primarily achieved in
going from the equator to about 45$^\circ$, with little further
decrease in $\Omega$ achieved at higher latitudes in most of the
cases.  Whereas the overall latitudinal contrasts from equator to pole
in the models and in the sun are roughly of the same order, the angular
velocity in the sun continues to slow down much more as the pole is
approached.  Two models, designated as {\sl LAM} (in Miesch et al. 2000) and {\sl L3} 
(in Elliott et al. 2000), do
exhibit $\Omega$ which decrease at high latitudes, but {\sl LAM}
involves an emerging heat flux that varies too much with latitude due
to choice of boundary conditions, and {\sl L3} has an overall $\Delta
\Omega$ that is only two-thirds of the helioseismic value.  Thus in
confronting {\sl Issue 1}, we will search in parameter space for solutions
that can achieve $\Omega$ profiles in which the decrease with latitude
does not taper off at mid latitudes and for which the contrast $\Delta
\Omega$ is at least comparable to the helioseismic findings.

As {\sl Issue 2}, with the convection becoming more turbulent, achieved
by decreasing either the thermal or viscous diffusivities, there is a
tendency for the latitudinal contrast $\Delta \Omega$ in the solutions
to diminish or even decrease very prominently, thus being at variance
with $\Delta \Omega$ deduced from helioseismology.  This behavior
appears to arise from increasing complexity leading to a weakening of
nonlinear velocity correlations that have a crucial role in angular
momentum redistribution.  These Reynolds stress terms are strong in the
laminar solutions that involve tilted columnar convection cells
(`banana cells') aligned with the rotation axis; they weaken as the
flows become more intricate, but would be expected to become again
significant once coherent structures develop at higher levels of
turbulence.  For example, the model {\sl TUR} (in Miesch et al. 2000) 
exhibits the emergence of
downflow networks involving fairly persistent plumes which possess some
of the expected attributes of the coherent structure seen in localized
domains of highly turbulent convection (e.g. Brummell et al. 1998).  As
a result, {\sl TUR} has a fairly interesting angular momentum transport
attributed to the nonlinear correlations that sustain a level of
differential rotation slightly weaker than {\sl LAM}, but it too has a
considerable variation of heat flux with latitude.  The model {\sl T2}
(in Elliott et al. 2000) 
sought to correct the latter by using modified thermal boundary
conditions but appears to not have attained high enough turbulence
levels to realize strong coherent structures.  Absent those features,
{\sl T2} yielded $\Omega$ profiles with a small $\Delta \Omega$, and
even a slightly slower equatorial rotation rate than that in the mid
latitudes.  Thus in confronting {\sl Issue 2}, we seek turbulent
solutions that possess $\Omega$ profiles with fast equators and strong
latitudinal contrasts $\Delta \Omega$, and emerging heat fluxes that
vary little with latitude.  To achieve this we have considered two
paths in parameter space that yield more turbulent solutions by either
varying the Prandtl number or keeping it fixed, while maintaining the
same rotational constraint as measured by a convective Rossby number.

We describe briefly in \S2 the ASH code and the set of parameters used
for the simulations studied here.  In \S3 we discuss the properties of
rotating turbulent convection and the resulting differential rotation
and the meridional circulation for the five cases $A$, $AB$, $B$, $C$
and $D$.  In \S4 we analyze the transport of angular momentum by
several processes and the influence of baroclinic effects in
establishing the mean flows.  In \S5 we reflect on the significance of
our findings, and especially in terms of dealing with the two issues raised
by the prior simulations with ASH.

\section{FORMULATING THE MODEL}

Our numerical models are intended to be a faithful if highly simplified
descriptions of the solar convection zone.  In brief overview, solar
values are taken for the heat flux, rotation rate, mass and radius, and
a perfect gas is assumed since the upper boundary of the shell lies
below the H and He ionization zones; contact is made with a real
solar structure model for the radial stratification being considered.
The computational domain extends from about 0.72$R_{\odot}$ to
0.96$R_{\odot}$, where $R_{\odot}$ is solar radius, with such shells
having an overall density contrast in radius of about 25, and as a
concequence compressibility effects are substantial.  Thus we are
concerned only with the central portion of the convection zone, dealing
neither with the penetrative convection below that zone, nor the two
shear layers present at the top and bottom of it.  Given the
computational resources available, we prefer to concentrate our effort
on processes that establish the primary differential rotation in the
bulk of the convection zone, and in future studies will seek to
incorporate the other regions.  We have as well softened the effects of
the very steep entropy gradient close to the surface that would
otherwise favor the driving of very small granular and mesogranular
scales of convection, with these requiring a spatial resolution at
least ten times greater than presently available.

The ASH code solves the 3--D anelastic equations of motion in a
rotating spherical shell geometry using a pseudo-spectral semi-implicit
approach (Clune et al. 1999).  As discussed in detail in Miesch et al.
(2000), these equations are fully nonlinear in velocity variables and
linearized in thermodynamic variables with respect to a spherically
symmetric mean state having a density $\rb$, pressure $\bar{P}$,
temperature $\tb$ and specific entropy $\sb$ and perturbations about
this mean state of $\rho$, $P$, $T$, $S$.  The conservation of mass,
momentum and energy (or entropy) in a rotating reference frame are thus
expressed as
\begin{equation}
\nab\cdot(\rb {\bf v}) = 0,
\end{equation}
\vspace{-0.5cm}
\begin{eqnarray}
\rb \left(\frac{\p {\bf v}}{\p t}+({\bf v}\cdot\nab){\bf v}+2{\bf
\Omega_o}\times{\bf v}\right)
 = -\nab P \nonumber \\ + \rho {\bf g} - \nab\cdot\mbox{\boldmath $\cal
D$}-[\nab\bar{P}-\rb{\bf g}],
\end{eqnarray}
\begin{eqnarray}
\rb \tb \frac{\p S}{\p t}&=&\nab\cdot[\kappa_r \rb c_p \nab
(\tb+T)+\kappa \rb \tb \nab (\sb+S)] \\ &-&\rb \tb{\bf v}\cdot\nab
(\sb+S)+2\rb\nu\left[e_{ij}e_{ij}-1/3(\nab\cdot{\bf v})^2\right]
\nonumber
\end{eqnarray}
where $c_p$ is the specific heat at constant pressure, ${\bf
v}=(v_r,v_{\theta},v_{\phi})$ is the local velocity in spherical
geometry in the rotating frame of constant angular velocity ${\bf
\Omega_o}$, ${\bf g}$ the gravitational acceleration, $\kappa_r$ the
radiative diffusivity, and ${\bf \cal D}$ the viscous stress tensor,
with components
\begin{eqnarray}
{\cal D}_{ij}=-2\rb\nu[e_{ij}-1/3(\nab\cdot{\bf v})\delta_{ij}],
\end{eqnarray}
where $e_{ij}$ is the strain rate tensor.  Here $\nu$ and $\kappa$ are
effective
eddy diffusivities for vorticity and entropy.  To close the set of equations,
the linearized relations for the thermodynamic fluctuations are
\begin{equation}\label{eos}
\frac{\rho}{\rb}=\frac{P}{\bar{P}}-\frac{T}{\tb}=\frac{P}{\gamma\bar{P}}
-\frac{S}{c_p},
\end{equation}
assuming the ideal gas law
\begin{equation}\label{eqn: gp}
\bar{P}={\cal R} \rb \tb ,
\end{equation}

\noindent where ${\cal R}$ is the gas constant.  The bracketed term in equation
(2), $\nab\bar{P}-\rb{\bf g}$, vanishes initially because the mean
state begins in hydrostatic balance from a one-dimensional radial solar
model (Brun, Turck-Chi\`eze \& Zahn 1999), but as the convection
becomes established this term becomes nonzero through effects of
turbulent pressure.  It is essential to take into account effects of
compressibility upon the convection, since the solar convection zone
spans many density scale heights.  To accommodate this, we use the
anelastic approximation (Gough 1969) to filter out the sound waves and
therefore permit bigger time steps for the temporal evolution.  The latter
is allowed since the CFL (Courant, Friedrichs \& Lewy) numerical
stability condition now applies to the smaller convective
velocities rather than the sound speed $c_s$.

Due to the small solar molecular viscosity, direct numerical
simulations (DNS) of the full scale range of motions present in stellar
convection zones are currently not feasible.  We seek to resolve the
largest scales of convective motion that we believe are the main
drivers of the solar differential rotation, doing so within a
large-eddy simulation (LES) formulation where $\nu$ and
$\kappa$ are assumed to be an effective eddy viscosity and eddy
diffusivity that represent unresolved subgrid-scale (SGS) processes,
chosen to suitably truncate the nonlinear energy cascade.  For
simplicity, both are here taken to be functions of radius alone, and
are chosen to scale as the inverse of mean density.  Other forms
that may be determined from the properties of the large-scale flows
according to one of many prescriptions (e.g. Lesieur 1997, Canuto 1999)
will be considered in the future.  We have also introduced an
unresolved enthalpy flux proportional to the mean entropy gradient in
equation (3) in order to account for transport by small-scale
convective structures near the top of our domain  (Miesch et al. 2000). 
We utilize the same radial profile for that mean eddy diffusivity in our five cases in
order to minimize the impact of our SGS treatment on the main
properties of our solutions. We emphasize that currently tractable simulations are still many
decades away in parameter space from the intensely turbulent conditions
encountered in the sun, and thus these large-eddy simulations must be
viewed as training tools for developing our dynamical intuition of what
might be proceeding within the solar convection zone.

Within the ASH code, the mass flux is imposed to be divergence-free by
using poloidal $W$ and toroidal $Z$ functions.  The thermodynamic variables
$P$ and $S$, and $W$ and $Z$, are expanded in spherical harmonics
$Y^m_{\ell}(\theta,\phi)$ to resolve their horizontal structures and in
Chebyshev polynomials $T_n (r)$ to resolve their radial structures.
This approach has the distinct advantage that the spatial resolution is
uniform everywhere on a sphere when a complete set of spherical
harmonics is used in degree $\ell$ (retaining all azimuthal orders
$m$).  We expand up to degree $\ell=\ell_{max}$ (depending on the
number of latitudinal mesh points $N_{\theta}$, e.g.
$\ell_{max}=(2N_{\theta}-1)/3$), utilize as longitudinal mesh points $N_{\phi}=2 N_{\theta}$,
and employ $N_r$ collocation points
in projecting upon the Chebyshev polynomials.  In this study the
highest resolution used has $\ell_{max}=340$ and $N_r=193$.  The time
evolution is carried out using an implicit, second-order
Crank-Nicholson scheme for the linear terms and an explicit,
second-order Adams-Bashforth scheme for the advective and Coriolis
terms.

Within ASH, all spectral transformations are applied to data local to
each processor, with inter-processor transposes performed when
necessary to arrange for the transformation dimension to be local.  The
triangular truncation in spectral space precludes any simple
distribution of the data and workload among the nodes. For very large
problems, the Legendre transformations dominate the workload and, as a
result, great care has been taken to optimize their performance on
cache-based architectures.  Arrays and loops have been structured to
operate on blocks which minimize cache misses.  The ASH code is
extremely flexible and has demonstrated excellent scalability on
massively parallel supercomputers such as the Cray T3E, IBM SP-3 and
Origin 2000.

As boundary conditions, we impose impenetrable and stress-free
conditions for the velocity field and constant flux (i.e constant
entropy gradient) both at the inner and outer boundaries.
We seek solutions with an emerging flux at the top that is invariant 
with latitude ({\sl Issue 2}). As initial conditions, we have started 
some simulations (cases $A$, $B$) from quiescent conditions
of uniform rotation, and others (cases $AB$, $C$, $D$) from evolved solutions in
which we modify certain diffusivities.  This leads to changes in the
effective Rayleigh number $R_a$, the Prandtl number $P_r$, the P\'eclet
number $P_e$, the Reynolds number $R_e$ and the Taylor number $ T_a$,
while keeping constant the convective Rossby number $R_{oc}$, all of
which are defined in Table 1.  We also summarize there the parameters
of the five simulation cases.

\section{PROPERTIES OF TURBULENT COMPRESSIBLE CONVECTION}

We have conducted five simulations involving increasingly nonlinear
flows that are achieved by reducing the viscous and entropy
diffusivities in the manner outlined in Table 1.  We have followed two
paths in parameter space in obtaining more complex convective flows.  
On {\sl Path 1} in going from case $A$ to $C$ via $B$, we
incrementally decreased the eddy viscosity $\nu$ while keeping the eddy
diffusivity $\kappa$ constant, thereby reducing the Prandtl number
$P_r$ by a factor 8.  In particular, the laminar case $A$ has $P_r$ of
unity; reducing the viscosity by a factor of 4 leads to the mildly
turbulent case $B$ with $P_r = 0.25$, or by a factor of 8 leads to the
more turbulent case $C$ with $P_r = 0.125$.  This serves to increase the
Reynolds number $R_e$ while only mildly increasing the P\'eclet number
$P_e$. {\sl Path 2} kept the Prandtl number fixed at $P_r = 0.25$ as the
complexity of the flows was increased by reducing both diffusivities.
Starting from case $AB$, we go to case $B$ by decreasing both
diffusivities $\nu$ and $\kappa$ by a factor of 2, and then to our most
turbulent case $D$ by further reducing both by a factor of 2 relative
to case $B$.  This {\sl Path 2} in going from case $AB$ to $D$ via $B$
results in both $R_e$ and $P_e$ increasing comparably.  All our models
possess a convective Rossby number $R_{oc}$ of order 2/3, thus
maintaining a strong rotational constraint on the convection.

As we shall describe in some detail, the resulting vigorous convection
influenced by rotation in all these cases is intricate and richly time
dependent, much as found in Miesch et al. (2000) and Elliott et al.
(2000). It is characterized by networks of strong downflow at the
periphery of the convection cells, and weaker upflows in their middle,
both of which are a consequence of the effects of compressibility as we
consider flows that can span multiple density scale heights in the
vertical.  Indeed, we consistently observe that the downflows are able to
extend over the full depth of the unstable layer, appearing as twisted
sheets of downflow near the top and more distinctive plumes deeper in
the layer.  These downflow networks essentially represent coherent
structures amidst the turbulence, and they are found to have a most
significant role in the nonlinear transport of angular momentum by
yielding correlations between different velocity components that form
Reynolds stress terms.  We find that the convection in all cases
studied here is able to redistribute angular momentum in such a manner
that substantial differential rotation profiles are established, the
properties of which are the major focus of this work.

\subsection{Complex evolution of convective patterns}

\begin{figure*}[!htp]
\centerline{\includegraphics[width=0.45\linewidth,angle=-90]{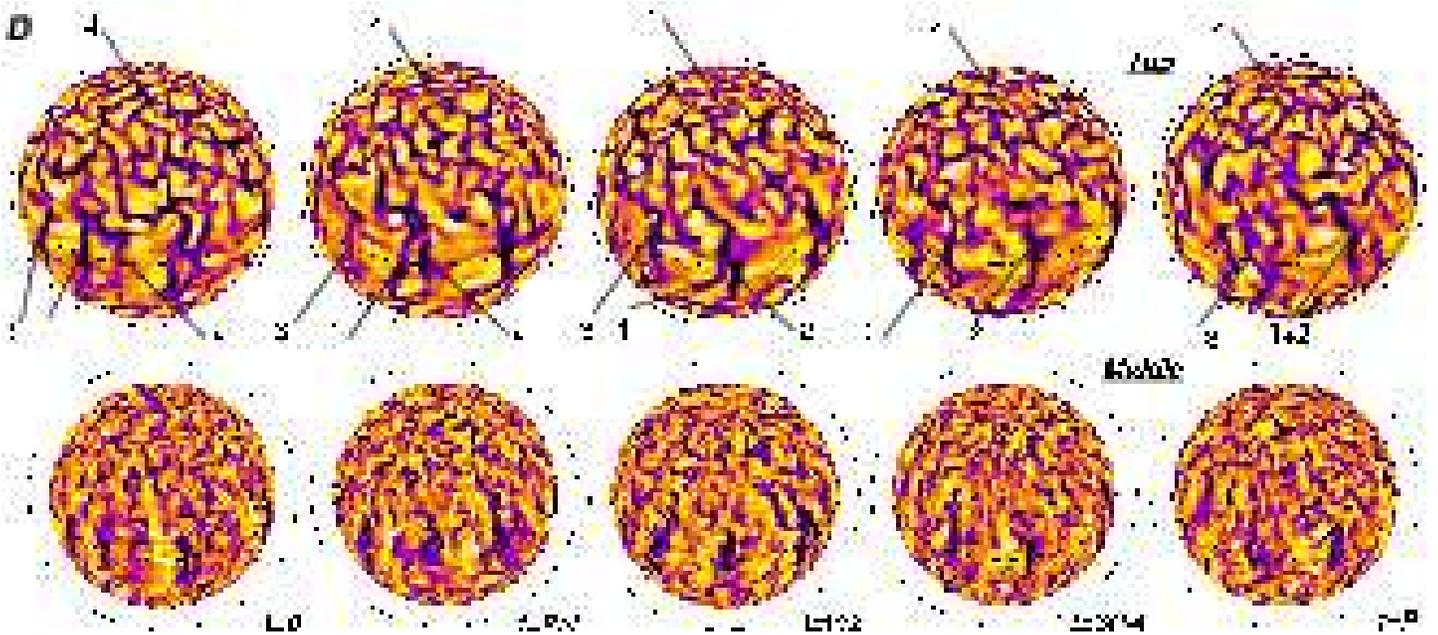}}

\caption[]{\label{fig2} Evolution of the convection throughout one solar
rotation, showing the radial velocity of case $D$ near the top and at
the middle of the domain. The time interval between each successive
image is about 7 days. Features {\sl 1} and {\sl 2} exhibit the merging of the downflow
lanes, feature {\sl 3} the shearing action of the differential rotation present in the
shell and feature {\sl 4} the appearance and deformation of a convective cell at higher
latitudes.}
\end{figure*}

The time dependence in our most turbulent simulation (case $D$) is
shown in Figure \ref{fig2} which displays two sequence of images of
the radial velocity on spherical surfaces over the course of one full
rotation.  The upper sequence with views near the top of the layer
involves simpler downflow networks (shown in darker tones) that are
easier to intercompare from frame to frame, whereas the lower ones with
views in the middle of the convecting layer are more difficult to track
because of increased complexity of the patterns in the more turbulent
flows there.  The vantage point is in the uniformly rotating frame used
in our modelling, and some of the pattern evolution results from the
prograde zonal flows at low latitudes and retrograde ones at high
latitudes associated with the differential rotation relative to this
frame. There is further melding and shearing of particular downflow
lanes as the convection cells evolve over a broad range of time scales,
some of which are comparable to the rotation period. This is
particularly evident in some of the downflow structures identified near
the equatorial region in the upper sequence, with features labeled {\sl 1}
and {\sl 2} illustrating the merging of two downflow lanes, and feature {\sl 3} the
typical distortion of a lane which also involves both a site of
cyclonic swirl in the northern hemisphere and another that is
appropriately anticyclonic in the southern hemisphere.  The behavior at
higher latitudes that involves retrograde displacement of the downflow
networks is somewhat more intricate, partly because the convection
cells are of smaller scale and exhibit the frequent formation of new
downflow lanes (as in feature {\sl 4}) that can serve to cleave existing
cells. Figure \ref{fig2} emphasizes that the overall pattern
of these global cells is sufficiently modified during the course of one
rotation period so that it would be difficult to identify particular
structures (relative to our uniformly rotating vantage point) when
viewed in a subsequent rotation. This would suggest that giant cells
possibly present within the solar convection zone may also loose their
identity from one Carrington rotation to the next.  This comes about
both because of advection and distortion of the cells by the mean zonal
flows associated with the differential rotation (here at the equator
leading to relative angular displacements in longitude of about
70$^{\circ}$ over one rotation period), and because of fairly rapid evolution and some
propagation in their individual downflow patterns.

\subsection{Downflow networks and variation with depth}

\begin{figure*}[btp]
\centerline{\includegraphics[width=0.95\linewidth]{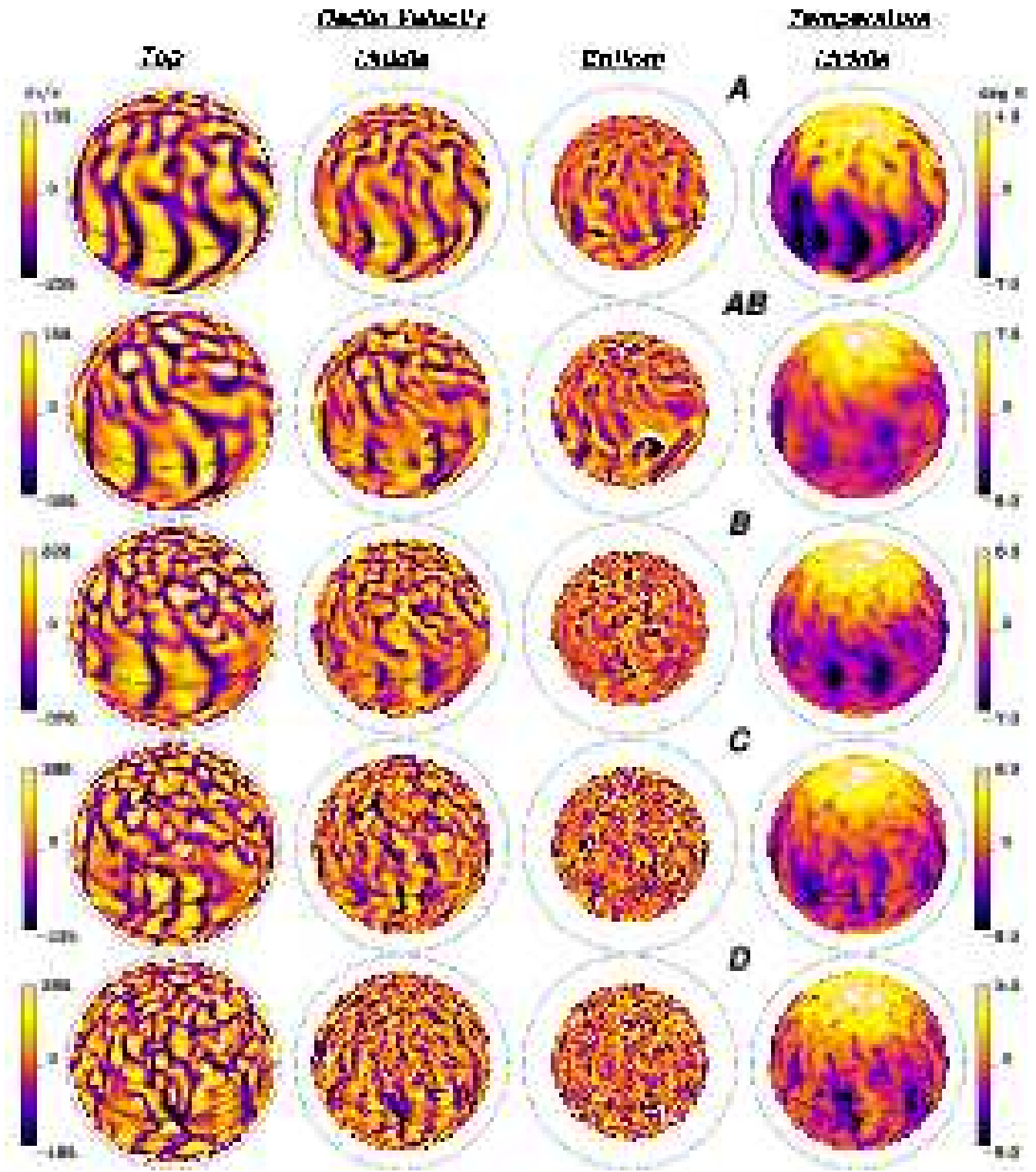}}

\caption[]{\label{fig3} Convective patterns for the five cases $A$, 
$AB$, $B$, $C$ and $D$ as increasingly turbulent flows are attained. The radial velocity shapshots are
 shown at three different depths (0.95, 0.84, 0.73
$R_{\odot}$). Downflows are represented in purple-dark tones and
upflows in orange-bright tones, with dynamic ranges indicated. The dotted circle is positioned at
radius R$_{\odot}$, and the equator is indicated by the dashed curve. The convective structures
become more complex in this progression of cases, with the banana-like convective
cells giving way to stronger and more frequent vortex sites. The strongest downflow lanes extend 
over the full depth range. The fluctuating temperature fields at mid depth are shown on the right,
 emphasizing that downflows are relatively cool and that the polar regions are on average warm.}
\end{figure*}

The convective structures as delineated by the downflow networks show
distinctive changes as the level of turbulence is increased in going
from case $A$ to case $D$. Figure \ref{fig3} provides an overview of
radial velocity snapshots in our five simulations at three depths (near
the top, middle and bottom), accompanied by the fluctuating temperature
fields at mid-depth. The upper surface in all our cases involves a
connected network of downflows surrounding broad upflows, but such
smoothness can disguise far more turbulent flows below.  The seemingly
cellular motions near the surface result from the expansion of fluid
elements rising through the rapidly decreasing density stratification
near the upper boundary, aided also by our increasing viscous and
thermal diffusivities there.  As viewed near the top, the tendency of
the convection in our laminar case $A$ to be organized into `banana
cells' nearly aligned with the rotation axis at low latitudes is
progressively disrupted by increasing the level of complexity in going
in turn to cases $AB$, $B$, $C$ and $D$. There is still some semblance
of north-south alignment in the downflows even in our most turbulent
case $D$, but the latitudinal span of this alignment is confined to a
narrow interval around the equator. Clearly the downflow lanes become
more wiggly and exhibit more pronounced vortical features and curvature
in this sequence of cases. As well, the downflow networks involve more
frequent branching points and smaller horizontal scales for the
convective patterns, especially at higher latitudes. Given the three
simultaneous views of the radial velocity, one can clearly identify
downflow lanes near the top in all our cases that turn into distinctive
plumes at greater depths, showing that organized flows extend over
multiple scale heights.  Indeed, the strongest downflows occur at the
interstices of the upper network and are able to pierce through the
interior turbulence, thus spanning the full depth range of the domain.

The plumes in the more turbulent cases $C$ and $D$ represent coherent
structures that are embedded within less ordered flows that surround
them.  They are able to maintain their identity, though with some
distortion and mobility, over significant intervals of time.  Although
these downflowing plumes are primarily directed radially inward, they
show some tilt both toward the rotation axis and out of the meridional
plane. This yields correlated velocity components and thus
Reynolds stresses that are a key ingredient in the redistribution of
angular momentum within the shell.  Such tilting away from the local
radial direction in coherent downflows has been seen in high-resolution
local $f$-plane simulations of rotating compressible convection
(Brummell et al. 1998), and their presence has a dominant role in
establishing the mean zonal and meridional flows. We also refer to 
Rieutord \& Zahn (1995) and Zahn (2000) for an analytical study of the 
transport properties and correlations present in such strong vortex structures
and on their potential dynamical role in the solar convection zone.

The strong downflows shown in Figure \ref{fig3} accentuate the
asymmetries that are characteristic of compressible convection, with
typical peak amplitudes in these downflows at mid-layer being as much
as twofold greater than that in the upflows.  As might be expected, the
overall rms radial velocities listed in Table 2 increase with
complexity in the flow fields in going from case $A$ to $D$.  The
asymmetries between upflows and downflows have the consequence that the
kinetic energy flux in such compressible convection is directed
radially inward, in contrast to the enthalpy and radiative fluxes that are
directed outward in transporting the solar flux (see Figure \ref{fig10}$a$).

That enthalpy flux involves correlations between radial velocities and
temperature fluctuations, and these are evidently strong as seen when
inspecting the temperature and velocity fields shown at mid layer in
Figure \ref{fig3}.  Buoyancy driving within our thermal convection
involves downflows that are cooler and thus denser and upflows that are
warmer and lighter than the mean; there are systematic asymmetries in
those temperature fluctuations, much as in the radial velocities.
Further, in comparing the temperature maps with those of radial
velocity in the middle of the layer, some of the temperature patterns
are evidently smoother, which is a consequence of the greater thermal
diffusivities than viscosities for cases with Prandtl numbers less than
unity.  A striking property shared by all these temperature fields is
that the polar regions are consistently warmer than the lower
latitudes, a feature that we will find to be consistent with a fast or
prograde equatorial rotation.

\subsection{Driving strong differential rotation}\label{sec diffrot}

\begin{figure*}[!htp]
\centerline{\includegraphics[width=0.93\linewidth]{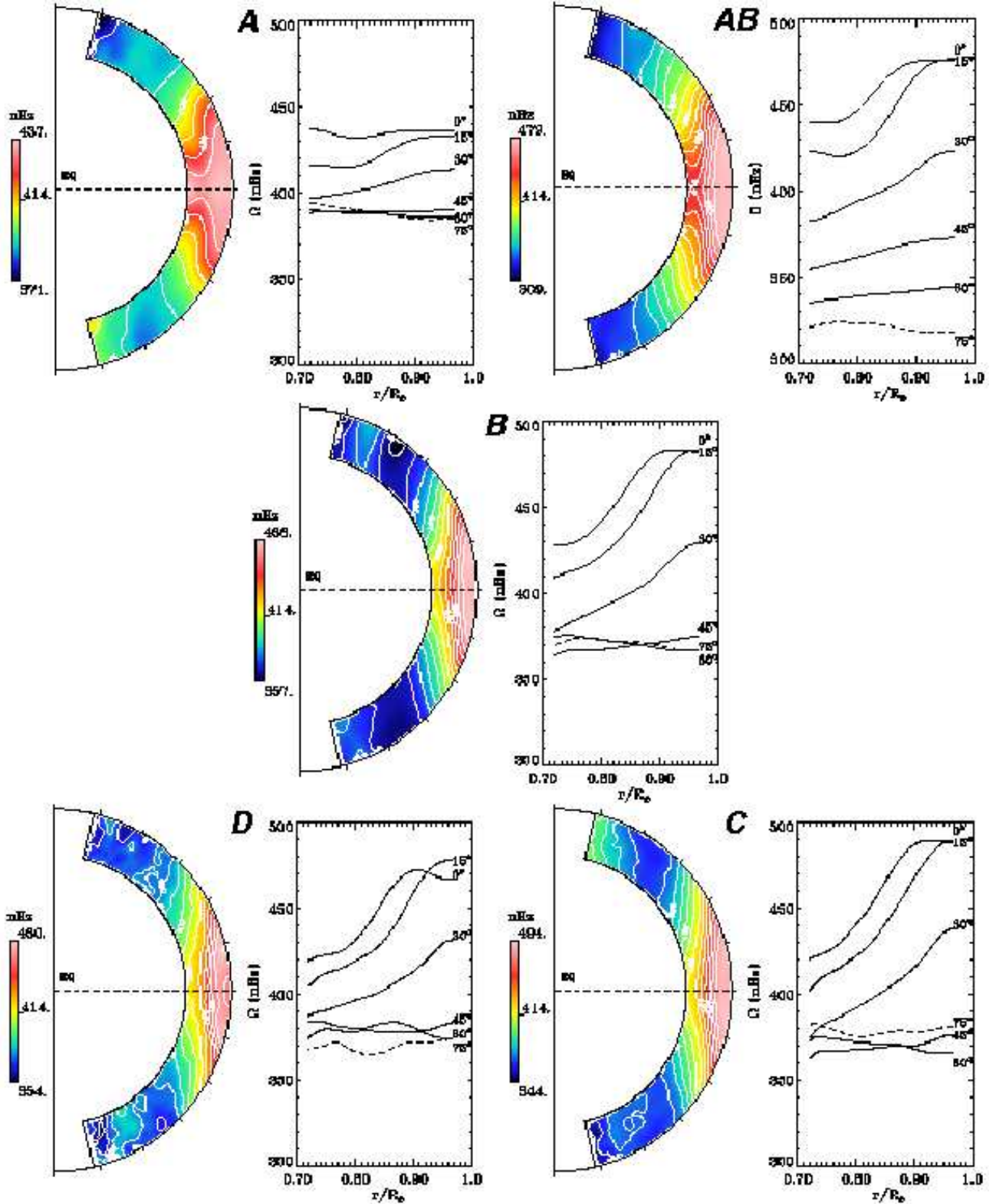}}
\vspace{0.0cm}
\caption[]{\label{fig4} Temporal and longitudinal averages of the angular velocity profiles
achieved in cases $A$, $AB$, $B$, $C$ and $D$, formed over intervals in turn 
 of 295, 275, 275, 175 and 35 days.
The contour plots for $\Omega/2\pi$ on the left of each panel are
independently scaled, whereas the radial cuts at the indicated
latitudes share the same scaling to accentuate the overall
behavior of the five cases.  The crossed layout of the five
cases emphasizes the two different paths followed to reach more 
turbulent states, mainly by lowering $P_r$ on {\sl Path 1} ($A \rightarrow B \rightarrow C$), 
and by lowering diffusivities while keeping $P_r$ constant on {\sl Path 2} ($AB \rightarrow B \rightarrow D$).
All cases exhibit a prograde
equatorial rotation and a strong contrast $\Delta \Omega$ from equator to pole.
Case $AB$ possesses a high latitude region of particularly slow rotation.}
\end{figure*}

The differential rotation profiles with radius and latitude that result
from the angular momentum redistribution by the vigorous convection in
our five simulations are presented in Figure \ref{fig4}. In order to
simplify comparison of our results with deductions drawn from
helioseismology (Fig.\ref{fig1}), we have converted our mean longitudinal velocities
$\hat{v}_{\phi}$ (with $\hat{}$ denoting averaging in longitude and
time) into a sidereal angular velocity $\Omega$ with radius and
latitude, and note that our reference frame rotation rate $\Omega_o
/2 \pi$ is 414 nHz (or a period of 28 days). The angular velocity in
all our cases exhibits substantial variations in time, and thus long
time averages must be formed to deduce the time mean profiles of
$\Omega$ shown in Figure \ref{fig4}.  The layout of the five cases in
Figure \ref{fig4} reflects the two paths we have taken in increasing
the complexity or turbulence level in the solutions: {\sl Path 1} in
going from $A$ to $C$ via $B$ while decreasing the Prandtl number takes
us from upper left to lower right, and {\sl Path 2} in going from
case $AB$ to $D$ via $B$ while keeping the Prandtl number fixed at
$P_r$=1/4 takes us from upper right to lower left.  Complexity in the
convective flows increases in going down the page.

All five simulations yield angular velocity $\Omega$ profiles that
involve fast (prograde) equatorial regions and slow (retrograde) high
latitude regions.  The variation of $\Omega$ with radius and latitude
may be best judged in the color contour plots in Figure \ref{fig4}
which are scaled independently for each of the cases; the reference
frame rate is also indicated.  The immediate polar regions are omitted
in these plots because it is difficult to obtain stable mean
$\Omega$ values at very high latitudes since the averaging domain there
becomes very small whereas the temporal fluctuations in the flows
remain substantial.  The contour plots reveal that there are some
differences in the $\Omega$ realized in the northern and southern
hemispheres, though such symmetry breaking is modest and probably will
diminish with longer averaging.  The convection itself is not
symmetric about the equator, and thus the mean zonal flows that
accompany such convection, and which are manifest as differential
rotation, can be expected to have variations in the two hemispheres.
In cases $B$, $C$ and $D$, there is some alignment of the
$\Omega$ contours at the lower latitudes with the rotation axis, thus
showing a tendency for $\Omega$ to be somewhat constant on cylinders.
Further, in these cases almost all the decrease in $\Omega$ with
latitude occurs in going from the equator to about $45^{\circ}$, or
thus is confined to the region outside the tangent cylinder to the inner
boundary (which intersects the outer boundary in our shell
configuration at $42^{\circ}$).  In contrast, cases $A$ and $AB$ show
far less alignment of $\Omega$ contours with cylinders at the lower
latitudes, and at mid-latitudes the contours are nearly aligned with
radial lines, more in the spirit of the helioseismic
inferences.  

Case $AB$ in Figure \ref{fig4} is unique in having the monotonic
decrease of $\Omega$ with latitude continue onward to high latitudes,
which is also the trend deduced from helioseismic measurements.  Thus
{\sl Issue 1} concerned with achieving a consistently decreasing $\Omega$
at high latitudes is resolved with case $AB$.  This is significant in
showing that such behavior can be realized in our modelling of convection
in deep shells, though it is not a common property in our other cases. 
It would be most desirable to understand how such high-latitude variation
in $\Omega$ is achieved in case $AB$, and we will address this in \S4.

The accompanying radial cuts of $\Omega$ at six fixed latitudes in
Figure \ref{fig4} permit us to readily quantify the $\Omega$ contrasts
with latitude achieved in these solutions, and to judge the functional
variation with radius in each case.  We use a common scaling for all
these line plots to make intercomparison between the cases most
convenient; the radial cuts for $\Omega$ have been averaged between the
north and south hemispheres.  Near the top of the convection zone at
radius 0.96$R$, the laminar convection in case $A$ produces a
differential rotation with a contrast in angular velocity, or $\Delta
\Omega / 2 \pi$, of about 50 nHz between the equator and 60$^{\circ}$,
or 12\% relative to the frame rotation rate (also quoted in Table 2).
Continuing on {\sl Path 1} in parameter space to the more turbulent
cases $B$ and $C$, we find that the latitudinal contrast in angular velocity has
increased substantially, becoming 115 nHz and 125
nHz in the two cases.  These correspond in turn to a 28\% and a 30\%
variation of the rotation rate.  These values are of interest since the
helioseismic inferences (Thompson et al. 1996, Schou et al. 1998, Howe
et al. 2000b) have a contrast of about 92 nHz at a similar depth between
the equator and 60$^{\circ}$, or a variation of about 22\% in rotation
rate, which further increases to about 32\% in going to 75$^{\circ}$.
The pronounced differential rotation in cases $B$ and $C$ is
accompanied by the $\Omega$ profiles becoming somewhat more aligned
with the rotation axis, resulting in steeper slopes in the radial cuts
at low and mid latitudes.  These two turbulent cases achieve their
larger $\Delta \Omega$ by both faster equatorial rotation rates and
slower rates at higher latitudes.  Thus {\sl Path 1} has been able
to resolve {\sl Issue 2}, concerned with retaining a strong contrast
$\Delta \Omega$ and a fast equator, as the solution becomes more
complex and turbulent.

Turning to {\sl Path 2} in parameter space, case $AB$ shows a contrast
of about 135 nHz between the equator and 60$^{\circ}$, or a 33\%
variation of rotation rate, which further increases to about 160 nHz or
39\% in going to 75$^{\circ}$.  The pivotal case $B$ has a somewhat
reduced contrast $\Delta \Omega/2\pi$ of 115 nHz or 28\% variation
between equator and 60$^{\circ}$, with little further variation at
higher latitudes.  The most turbulent case $D$ has a $\Delta \Omega/2\pi$ of
about 105 nHz or a 25\% variation between the equator and
60$^{\circ}$.  Thus {\sl Path 2} leads to a slight reduction in $\Delta
\Omega$ with increasing complexity, unlike the behavior of {\sl Path
1}.  However, even this path yields a turbulent solution case $D$ whose
$\Delta \Omega$ is still close to the helioseismic contrast, thus
largely resolving {\sl Issue 2}.  This is reemphasized in Figure
\ref{fig5} which summarizes the variation of $\Delta \Omega$ with $P_r$
for our five cases.

\vspace{0.2cm}
\centerline{\includegraphics[width=0.65\linewidth,angle=90]{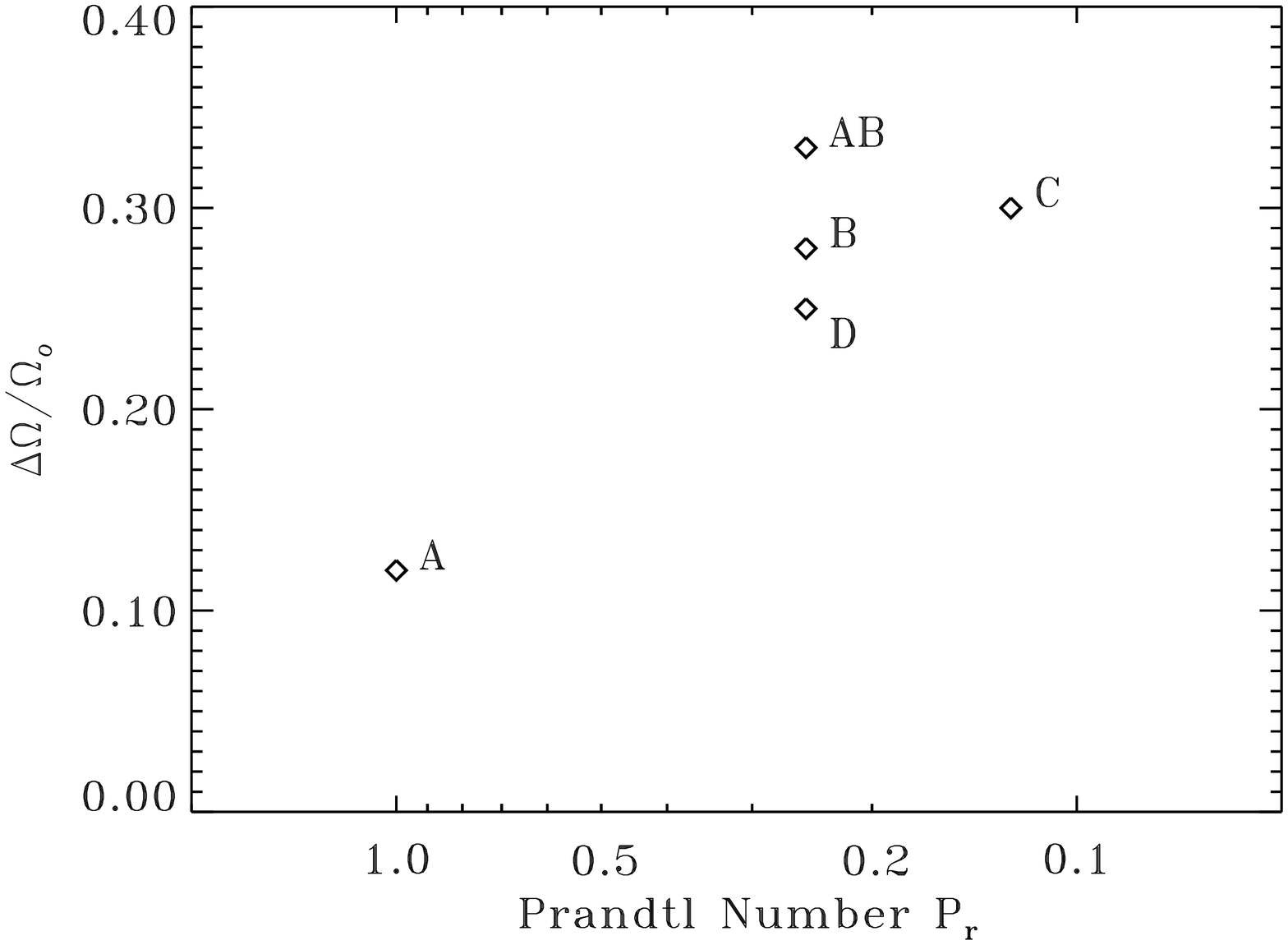}}

\figcaption[]{\label{fig5} Parameter space diagram for relative latitudinal 
angular velocity contrast $\Delta\Omega/\Omega_o$ as a function of the
Prandtl number $P_r$ for the five cases. The two paths toward
higher levels of turbulence either reduce $P_r$ ($A \rightarrow B \rightarrow C$), 
or maintain a constant $P_r$ ($AB \rightarrow B \rightarrow D$).}
\vspace{0.2cm}

Most of our cases possess overall latitudinal contrasts in $\Omega$
that are in the realm of solar values deduced from inversion of
helioseismic data, yet case $AB$ stands out in having the systematic
decrease of $\Omega$ with latitude extending almost to the poles, which
appears to be another distinguishing feature of the actual solar
$\Omega$ profiles. Further, case $AB$ displays little radial variation
in $\Omega$ at intermediate and high latitudes (from say 45$^{\circ}$
onward) as the angular velocity continues to decrease poleward.  Such
behavior is most interesting, and it is necessary to understand just
which convective properties within case $AB$ allow it to come into
reasonable contact with the helioseismic profiles for $\Omega$ deduced
in the bulk of the solar convection zone.

The $\Omega$ profiles in Figure \ref{fig4} have been formed from temporal
averages spanning multiple rotation periods as indicated.  It is
appropriate to consider if these represent truly `spun-up' solutions in
a statistical sense, and further whether several distinctive $\Omega$
profiles could be achieved for the same control parameters.  Both
issues may be intertwined, for the rate of approach to equilibration
can be influenced by the attraction characteristics of that
differential rotation state, and of course by the amplitude of the
fluxes available to redistribute angular momentum to achieve that state
(see \S4.1).  This overall dynamical system of turbulent convection is
sufficiently complex that we are uncertain whether there may exist
multiple basins of attraction leading to different classes of
differential rotation.  For instance, is the behavior of case $AB$ with
noticeably slow rotation at high latitudes an example of one class of
behavior, and our other cases that of another family?  Could such
families overlap in parameter space, or are there just gradual
variations in behavior in $\Omega$ with changes in the parameters?  We
have so far sought to address some of these questions by perturbing the
evolving solutions to see if they might flip to another state, but they
have not done so.  We plan to examine such issues of solution
uniqueness in our follow on studies in which we seek to extend the
slow-pole characteristics of case $AB$ to other parameter settings
involving more complex convection.

As to the relative maturity of the spun-up states shown in Figure \ref{fig4},
these vary from case to case due to the rapidly increasing
computational expense in dealing with the finer spatial resolution
required by the more complex simulations.  Cases $AB$ and $C$ were both
started from case $B$ that had already been run for over 4000 days of
elapsed simulation time (or a nominal 143 rotation periods involving
about 28 days each).  At this point case $B$ appeared to be
statistically stationary in terms of the kinetic energy associated with
the differential rotation, though it like most of the other simulations
exhibit small fluctuations in $\Omega$ profiles determined from
single-rotation averages, especially at the higher latitudes.  Case
$AB$ was evolved for about 2300 days (82 rotation periods), and we
illustrate in Figure \ref{fig6}$a$ a succession of $\Omega$ profiles with
latitude sampling the last 600 days in the simulation.  The solid curve
there is an average formed over 10 rotations (consistent with the
contour plot in Figure \ref{fig4}), and we see that the individual rotation
averages being sampled form a narrow envelope around it.  There is
evidently some symmetry breaking between the two hemispheres.  We
believe that the differential rotation for case $AB$ is now an
effectively stationary state (as is also confirmed in studying the
angular momentum flux balance in Figure \ref{fig11}).  The more turbulent case
$C$ was evolved for about 500 days (18 rotations) after being initiated
from case $B$, and a set of its angular velocity profiles are shown
sampling the last 300 days in Figure \ref{fig6}$b$.  We are less certain of its
stationarity, but we could not detect any systematic trends in the
evolution of its differential rotation over the last 10 rotations.  We
saw no evidence of a slow pole developing, but that may well require
more extended computations than could be presently arranged.  Figure
\ref{fig6} serves to emphasize that the angular velocity even in the sun may
be expected to vary somewhat from one rotation to another, with the
samplings here providing a sense of the amplitude of those changes.

\vspace{0.2cm}
\centerline{\includegraphics[width=0.8\linewidth]{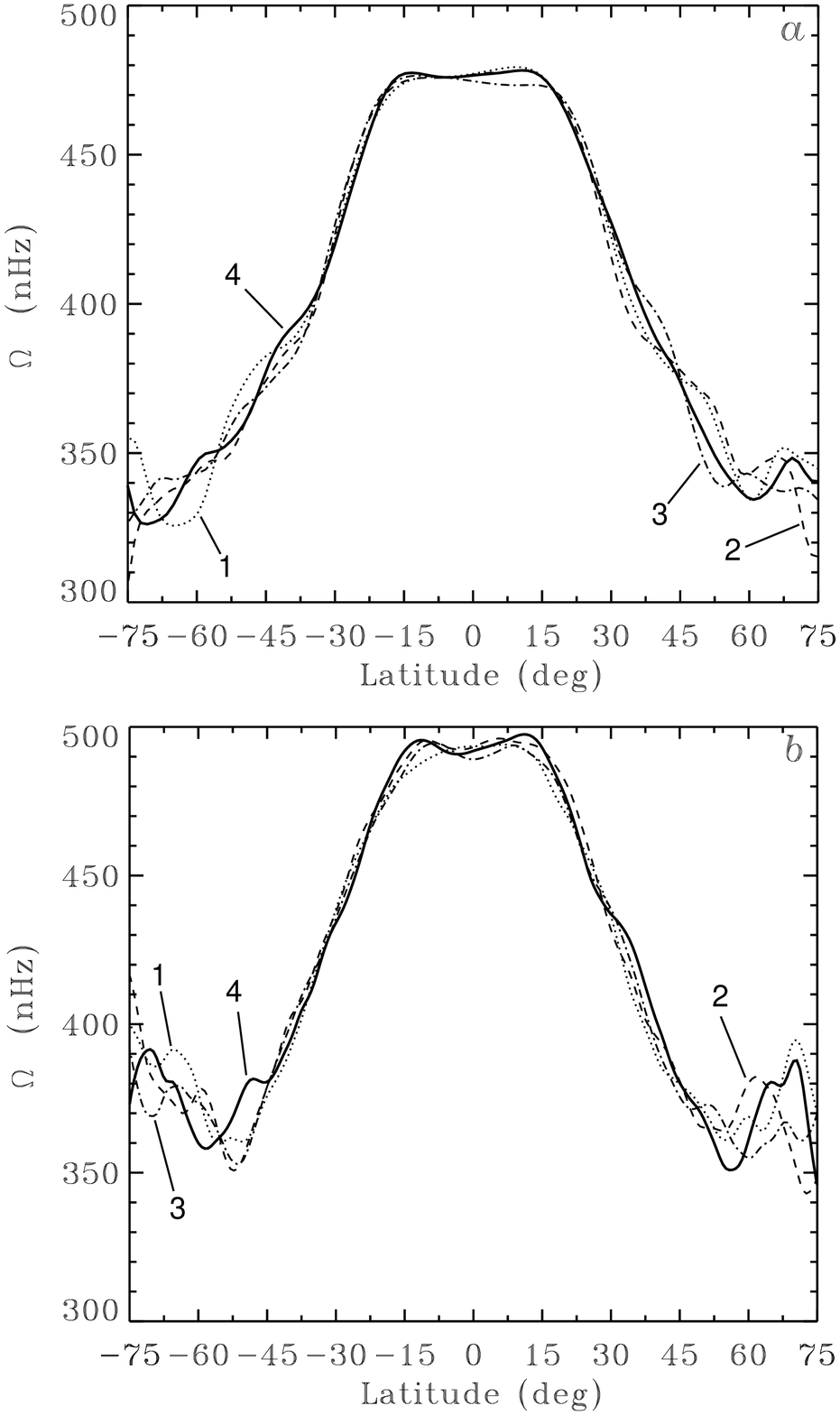}}
\figcaption[]{\label{fig6} Succession of time-averaged $\Omega$ profiles with latitude
at $r = 0.96 R_{\odot}$.  ~$(a)$ For case $AB$, a numbered sequence of
single-rotation averages spanning an interval of 600 days in the late
evolution of the system, with the bold curve {\sl 4} denoting an average over
the last 275 days in the simulation. ~$(b)$ For case $C$, dealing with
samples in a 300 day interval, and the bold curve {\sl 4} representing an
average over the final 175 days.  The variations are representative
of small changes in the differential rotation that accompany changes
in the convection once a mature statistical state has been achieved.}
\vspace{0.2cm}

\begin{figure*}[!ht]
\centerline{\includegraphics[angle=-90,width=1.0\linewidth]{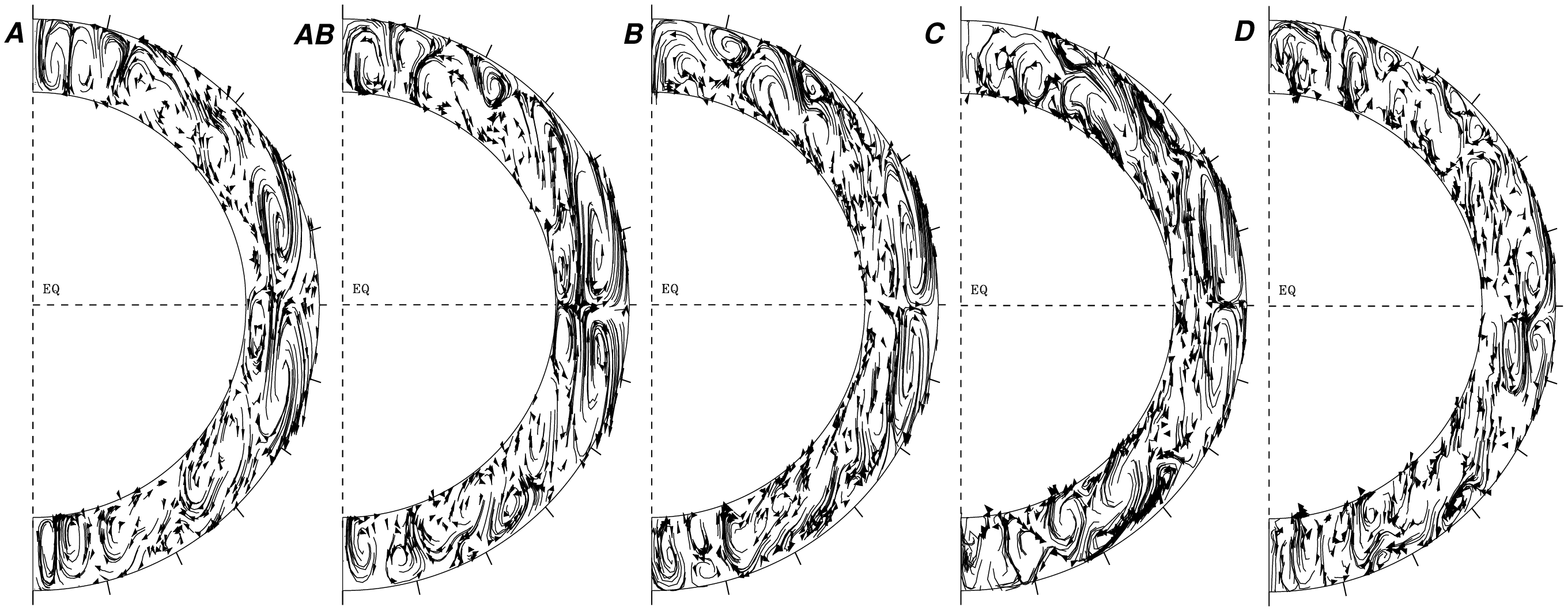}}
\caption[]{\label{fig7} Temporal and longitudinal averages of the meridional flows achieved in the cases $A$, $AB$, $B$, $C$ and $D$, deduced from sampling in turn 295, 275, 275, 175 and 35 days. Shown are random streaklines whose length is proportional to flow speed, with arrowheads indicating flow sense. The typical speeds in these meridional circulations are about 20 m$\,$s$^{-1}$. For all the cases strong poleward cells are present near the surface at low latitudes as well as 
return flows at mid depth.}
\end{figure*}

\subsection{Meridional circulation patterns}\label{sec mcirc}

The time-averaged meridional circulations that accompany the vigorous
convection in the five cases are shown in Figure \ref{fig7}.  The
typical amplitudes in these large-scale circulations are about 20
m\,s$^{-1}$, and thus comparable to the values deduced from local
domain helioseismic probing of the uppermost convection zone based
either on time-distance methods (e.g. Giles, Duvall \& Scherrer 1998) or
ring-diagram analyses (Schou \& Bogart 1998, Haber et al. 1998).  There is little change in
meridional circulation amplitudes as we increase the level of
turbulence in going from case $A$ to $D$.  However, multi-cell
structures in these circulations become more intricate with the increased
complexity of the convection.  At lower latitudes the circulation in
both hemispheres is poleward near the top of the domain, with return
flows at various depths. All cases display multiple cells with radius and
latitude, and never only one big meridional cell as often used in  mean field
models dealing with differential rotation (e.g. Rekowski \& R\"udiger 1998,
Durney 2000) or with Babcock-Leighton dynamos (e.g. Choudhuri, Sch\"ussler \& Dikpati 1995, Dikpati \& Charbonneau 1999). 
The resulting axisymmetric meridional circulation is maintained by
Coriolis forces acting on the mean zonal flows that appear as the
differential rotation, by buoyancy forces, by Reynolds stresses, and by
pressure gradients.  Given these competing processes, it is not self
evident as to what pattern of circulation cells should result, nor how
many should be present in depth or latitude.  Our five simulations have
shown that there is some variety in the meridional circulations
achieved, all of which involve multi-celled structures.  Since the
kinetic energy in the meridional circulation (MCKE) is typically about
two orders of magnitude smaller than in the differential rotation
(DRKE), as we will detail in \S3.5 and in Table $2$, small
variations in the differential rotation can yield substantial changes
in the circulations.  This is likewise true of the time-varying
Reynolds stresses from the evolving convection which again has a
kinetic energy (CKE) much larger than that of the meridional
circulations.  This may explain the complex time dependence realized by
the meridional flows, and the need to use long time averages in
defining their mean properties.

Another rendition of the time-averaged meridional circulations achieved
in cases $AB$ and $C$ is shown in Figure \ref{fig8} using a streamfunction
$\Psi$ based on the zonally-averaged mass flux [as in Miesch et al.
2000, equation (7)].  In case $AB$ (Fig. \ref{fig8}$a$) there are two
circulation cells positioned above each other in radius at low
latitudes.  The stronger upper one (solid contours representing
counterclockwise circulation) involves poleward flow that extends from
the equator to about 30$^{\circ}$ latitude near the top of the domain
in the northern hemisphere.  The southern hemisphere has likewise
poleward flow near the top at low latitudes, with ascending motions
again present from the equator to about 20$^{\circ}$ latitude.  At
latitudes greater than about 30$^{\circ}$ the relatively weak flow near
the top is mainly equatorward in both hemispheres, but exhibits
fluctuations. 

A quantitative measure of this for case $AB$ is provided
in Figure \ref{fig9}$a$ that displays the mean velocity component $\hat
v_{\theta}$ with latitude at two depths near the top of the domain.
The poleward flow in both hemispheres peaks at about 20$^{\circ}$
latitude and then decreases rapidly, changing to weak equatorward flow
above 30$^{\circ}$ which attains about one-third that peak amplitude. 

\vspace{0.2cm}
\centerline{\includegraphics[width=0.95\linewidth]{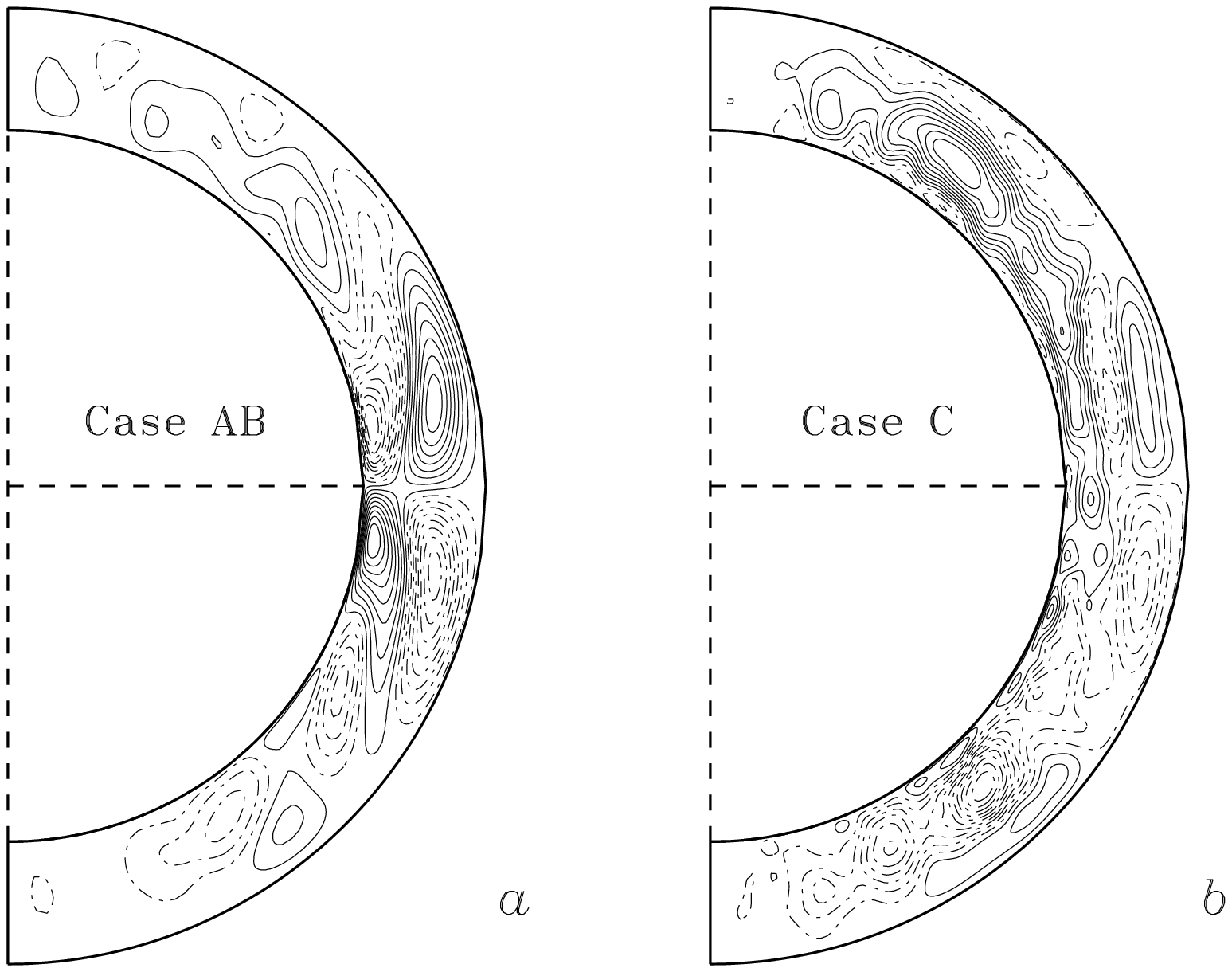}}
\figcaption[]{\label{fig8} Streamlines of the mean axisymmetric meridional circulation achieved in $(a)$ case $AB$ averaged over 275 days, and in $(b)$ case
$C$ averaged over 175 days.  Solid contours denote counterclockwise
circulation (and dashed clockwise), equally spaced in value. In case $AB$,
two circulation cells are present with radius at low latitudes, and only weak circulations at latitudes above 30$^{\circ}$. Case $C$ possesses three cells at low latitudes, with the deepest extending prominently to high latitudes.}
\vspace{0.2cm}

Turning to case $C$ in Figure \ref{fig8}$b$, it exhibits three circulation cells
positioned radially at low latitudes, with the outermost again yielding
poleward flow at the top of the domain that extends to about
35$^{\circ}$ in latitude.  At higher latitudes the mean meridional flow
is again equatorward near the top, attaining a peak amplitude for $\hat
v_{\theta}$ (detailed in Figure \ref{fig9}$b$) that is comparable to the
poleward one from the low latitudes, unlike in case $AB$.  Of the three
meridional cells at low latitudes in Case $C$, much as for model {\sl
TUR} in Miesch et al. 2000 (cf. Fig. $16a$), the deepest cell involves
a strong counterclockwise circulation that extends to high latitudes,
yielding a submerged poleward flow there that lies below the
equatorward flow at the top of the domain.  Such behavior involving a
third deep circulation cell that extends to high latitudes is also seen
in cases $B$ and $D$.  Such a strong third cell appears to be of
significance in the continuing net poleward transport of angular
momentum by the meridional circulations (cf. \S4.1, Fig. \ref{fig11}) in all
these cases at latitudes above about 30$^{\circ}$.  This is not
realized in case $AB$, and may contribute to its slow pole behavior.
It is encouraging that we have poleward circulations in the upper
regions of the simulations, which is in accord with the general sense
of the mean flows near the surface being deduced from local
helioseismology, though two cell behavior with latitude has been
detected recently only in the northern hemisphere near the peak of
solar activity (Haber et al. 2000). Such symmetry breaking in the two
solar hemispheres is an interesting property, and one that is also
occasionally realized in our simulations as the convection patterns evolve.
The helioseismic probing with ring diagram methods and explicit inversions 
is able to sense the meridional circulations only fairly close to the solar
surface, typically extending to depths of about 20 Mm or to radius 0.97
$R_{\odot}$, whereas our simulations have their upper boundary slightly below
this level at 0.96 $R_{\odot}$.

\vspace{0.2cm}
\centerline{\includegraphics[angle=90,width=0.9\linewidth]
{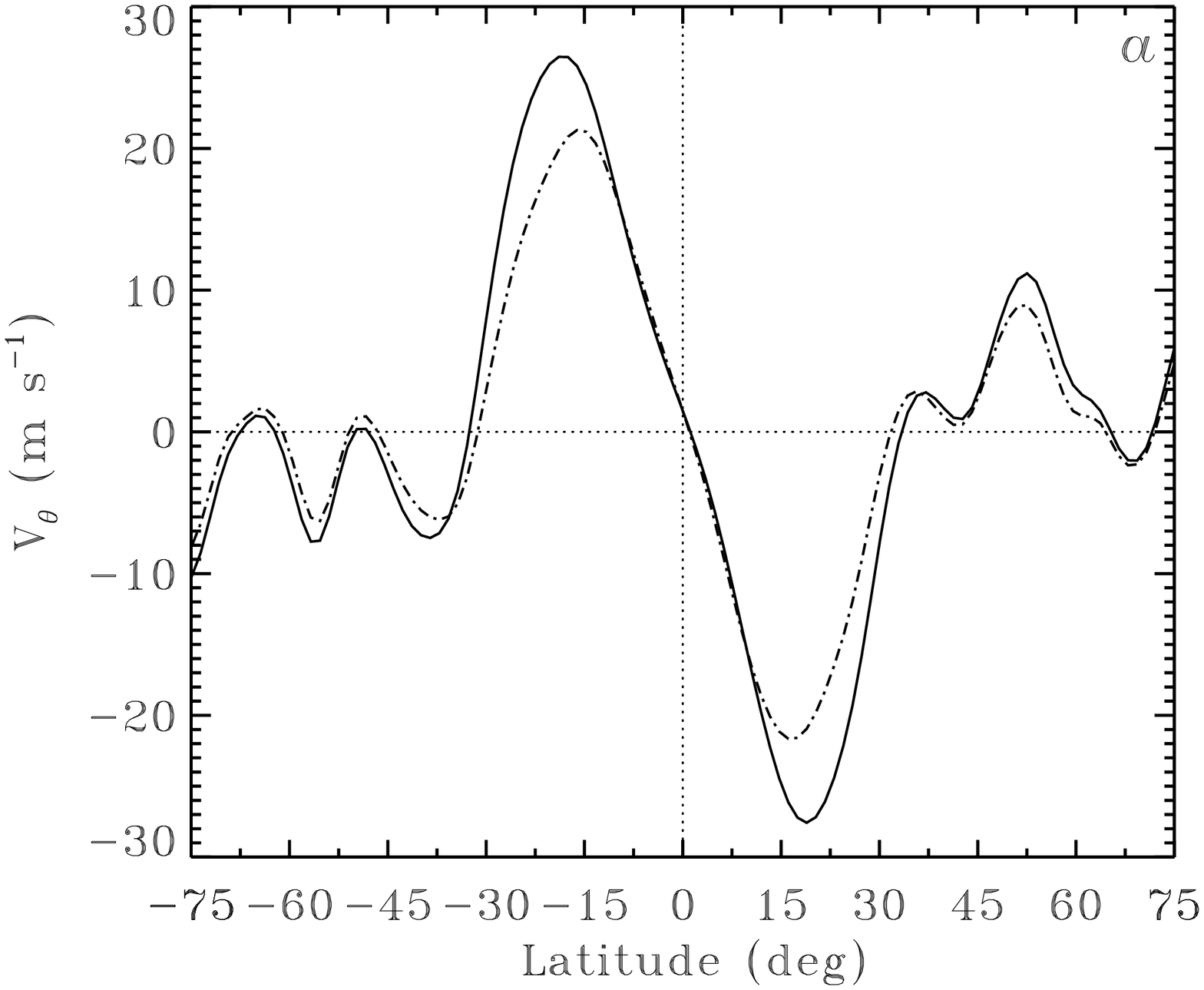}}
\vspace{0.5cm}
\centerline{\includegraphics[angle=90,width=0.9\linewidth]
{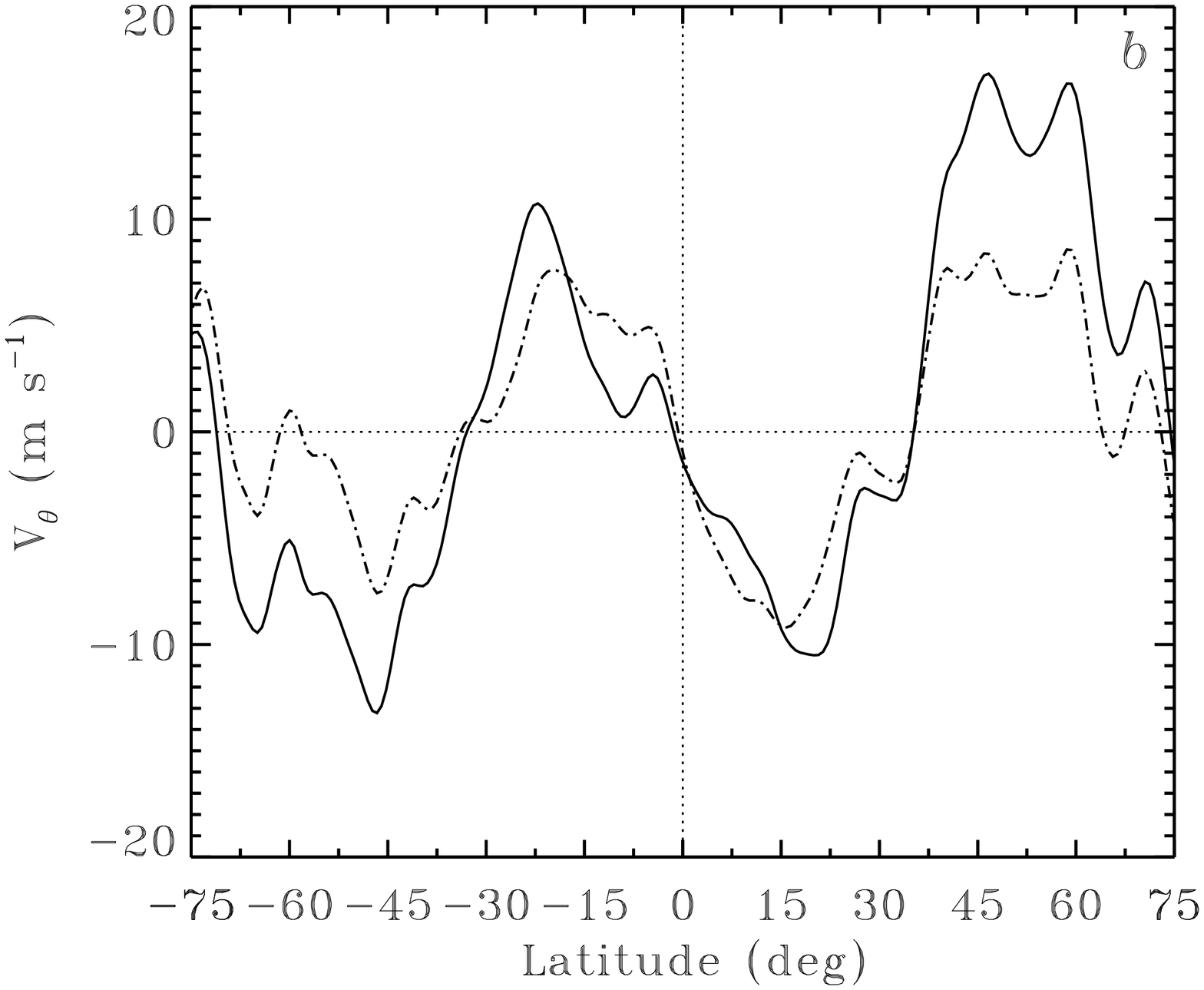}}
\vspace{0.2cm}
\figcaption[]{\label{fig9} Mean velocity component $\hat v_{\theta}$ with latitude at the two depths $r = 0.96 R_{\odot}$ ({\it solid}) and $0.94 R_{\odot}$
({\it dash dot}), showing in $(a)$ case $AB$ and in $(b)$ case $C$.
Positive values correspond to flow directed from north (positive
latitudes) to south (negative latitudes). At low latitudes the flows are poleward in both hemispheres, but whereas case $C$ exhibits fairly strong equatorward flow at latitudes above 35$^{\circ}$, case $AB$ possesses much weaker circulations there.}
\vspace{0.2cm}

  Thus we must be cautious in interpreting
similar behavior in the meridional circulations since our models and
the ring diagram analysis do not explicitly overlap in radius. 
Helioseismic assessments based on time-distance methods (Giles 1999,
Chou \& Dai 2001) and annular rings centered on the poles (Braun \& Fan
1998) report detecting effects attributable to meridional circulations
with a mainly poleward sense to depths corresponding to 0.90
$R_{\odot}$ or even 0.85 $R_{\odot}$.  Such results are most
interesting, but considerable further work on inversions would be
required to provide detailed profiles of the circulations with depth.
As these mappings become available, they may be able to confirm or
refute the multi-cell radial structure of meridional circulation (Fig.
\ref{fig7}) typically realized in our simulations.

\subsection{Energetics of the convection and the mean flows}

The overall energetics within these shells of rotating convection have
some interesting properties, in addition to the mean zonal and
meridional flows that coexist with the complex convective motions. The 
convection is responsible for transporting outward the solar flux
emerging from the deep interior.  We should recall, as discussed in
detail in Miesch et al. (2000), that the radial flux balance in these
convective shells involves four dominant contributors, namely the enthalpy or
convective flux $F_e$, the radiative flux $F_r$, the kinetic energy
flux $F_k$, and finally the unresolved eddy flux $F_u$, which add up to
form the total flux $F_t$. Figure \ref{fig10}$a$ shows the flux balance
with radius achieved in our most turbulent case $D$ as averaged over
horizontal surfaces and converted to luminosities.  The radiative flux
becomes significant deep in the layer due to the steady increase of
radiative conductivity with depth, and indeed by construction it
suffices to carry all the imposed flux through the lower boundary of
our domain where the radial velocities and thus the convective flux
vanishes.  A similar role near the top of the layer is played by the
sub-grid-scale turbulence that yields $F_u$, which being proportional
to a specified eddy diffusivity function $\kappa$ and the mean radial
gradient of entropy, suffices to carry the total flux through the upper
boundary and prevents the entropy gradient there from becoming too
superadiabatic compared to the scales of convection that we are
prepared to resolve spatially.  Over most of the interior of the shell,
the strong correlations between radial velocities and temperature
fluctuations yield the enthalpy flux $F_e$ that transports upward
almost all of the imposed flux, and this peaks near the middle of the
layer.  The kinetic energy flux $F_k$ works against the others by being
directed downward, a result of the fast downflow sheets and plumes
achieved by effects of compressibility (Hurlburt, Toomre \& Massaguer
1986).  These general properties are shared by our five cases, all of
which have achieved good overall flux balance with radius, as can be
assessed by examining $F_t$.

Figure \ref{fig10}$b$ presents the kinetic energy spectra with azimuthal
wavenumber $m$ at three depths, and averaged in time, as realized in
the case $D$ simulation.  The spectra are fairly broad, with a plateau
of power extending up to about $m$=30 corresponding to some of the most
vigorously driven scales, and then a rapid decrease involving about 5
orders of magnitude to the highest wavenumber of 340.  The decrease is
more rapid for the spectra formed near the top of the shell.  These
spectra suggest that the flows are well resolved, with a reasonable
scale separation between the dominant energy input range and the wide
interval over which dissipation functions.  We cannot readily identify
a clear inertial subrange, though for reference we include some power laws.
We also refer to Hathaway et al. (1996, 2000) for a discussion
of recent observational inferences about the solar kinetic energy
spectrum which does not seem to indicate any clear scaling law.

\vspace{0.2cm}
\centerline{\includegraphics[angle=90,width=0.95\linewidth]
{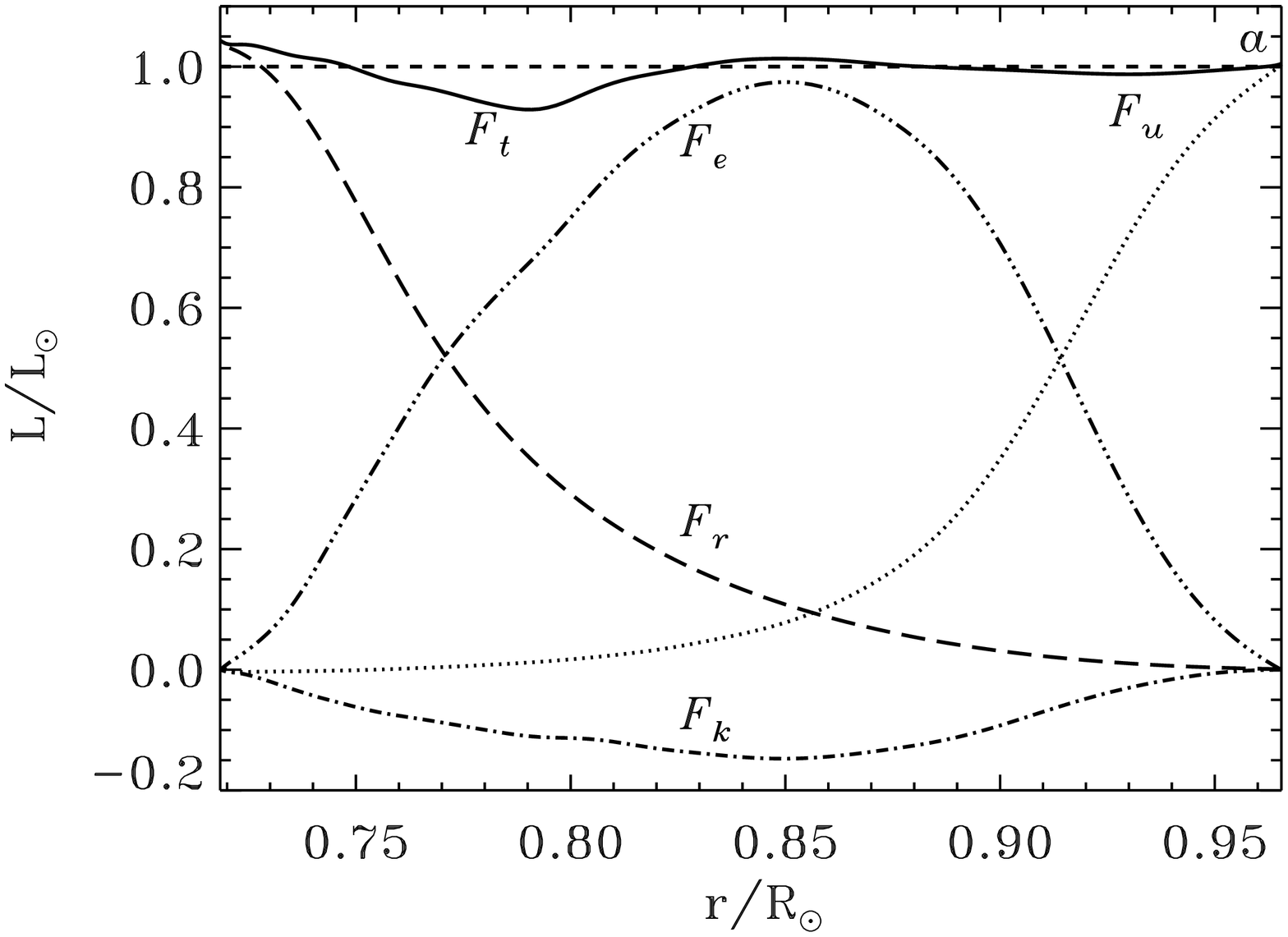}}
\centerline{\includegraphics[angle=90,width=0.95\linewidth]
{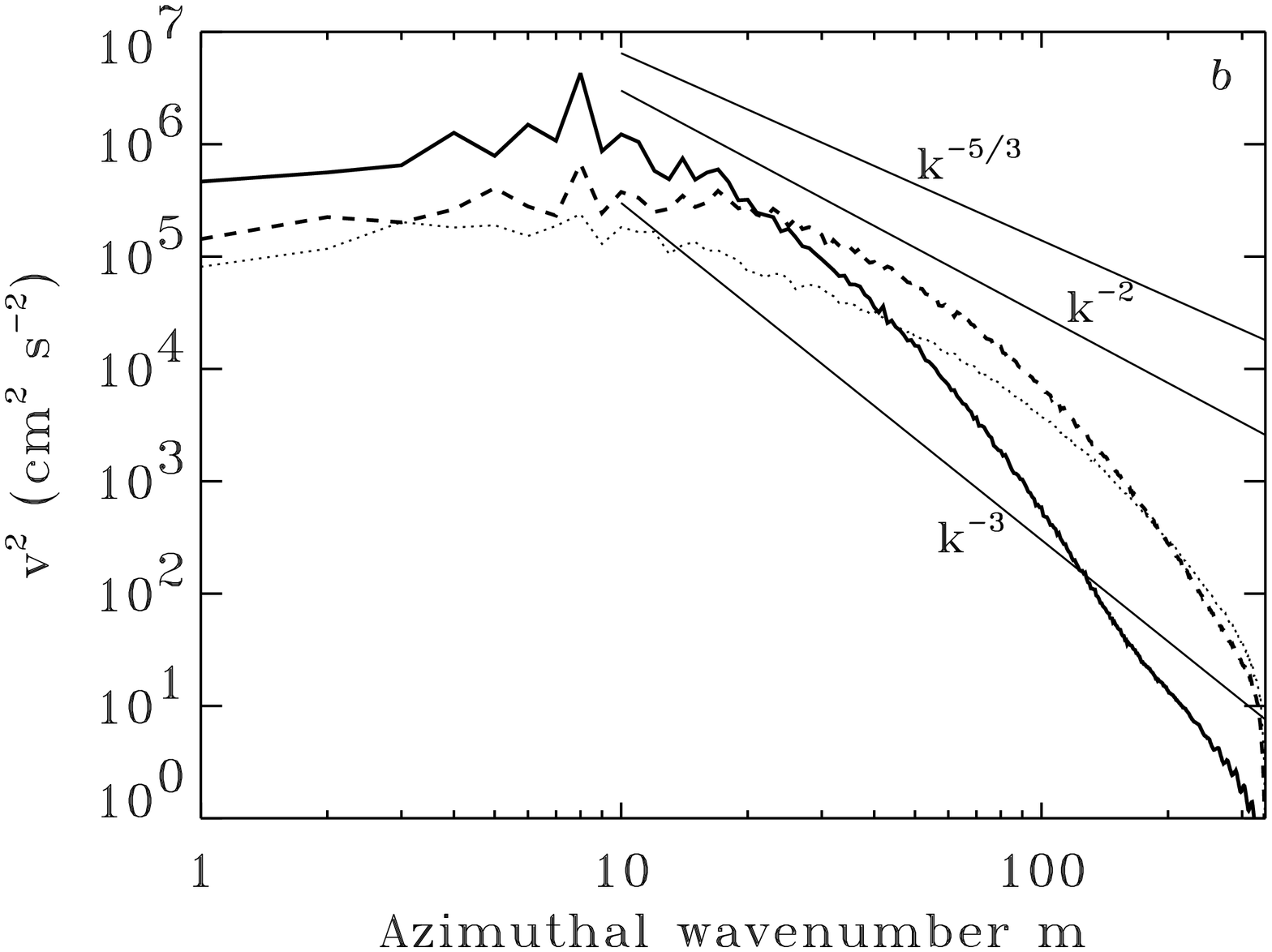}}
\vspace{0.0cm}
\figcaption[]{\label{fig10} ($a$) Radial transport of energy in case $D$ achieved by the fluxes
 $F_{r}$, $F_{e}$, $F_{k}$ and $F_{u}$, and their total $F_t$, all normalized by the 
 solar luminosity. 
($b$) Time averaged azimuthal wavenumber spectra of the kinetic energy on different 
radial surfaces ({\it solid} curve $r/R_{\odot}$=0.95; {\it dashed}, 0.84; {\it dotted}, 0.73) 
for case $D$ with $l_{max}=340$. The results have been averaged over 35 days. 
Superimposed are the power laws $k^{-5/3}$, $k^{-2}$ and $k^{-3}$;
no clear inertial subrange is identifiable.}
\vspace{0.2cm}

Table 2 summarizes various rms velocities that characterize our five
simulations as sampled in the middle of the layer where the enthapy
flux also peaks.  The rms radial velocity $\tilde v_r$ increases
monotonically in going through the sequence of cases $A$, $AB$, $B$,
$C$ to $D$.  The associated rms Reynolds number $\tilde R_e$ in Table 1
increases also (though part of this is due to changes in the
diffusivities), varying by a factor of about 15 from our laminar to
most turbulent solutions.  The rms longitudinal velocity $\tilde
v_{\phi}$ has the greatest amplitude in all the cases. However a
removal of the mean zonal flow component responsible for the
differential rotation yields $\tilde v_{\phi}'$. Comparison of this
with the radial and latitudinal rms velocities reveals that all possess
very comparable amplitudes, suggesting fairly isotropic convective
motions near the midplane.  Table 2 also assesses the
volume and time averaged total kinetic energy (KE), that associated
with the differential rotation (DRKE), with the meridional circulation
(MCKE), and with the convection itself (CKE).  In all of our solutions
the DRKE and CKE are comparable, and the MCKE is much smaller.  Table 2
reveals a most interesting contrast in behavior for the two paths.
Following {\sl Path 1} (involving $A$, $B$ and $C$, with decreasing
$P_r$), we find that KE in the solutions increases steadily with
increasing flow complexity.  This would be expected since the buoyancy
driving has strengthened relative to the dissipative mechanisms as
measured by the increasing Rayleigh number $R_a$ (Table 1).  {\sl Path 2}
(involving $AB$, $B$, and $D$, with $P_r$ kept fixed at 0.25) is quite
different as $R_a$ increases: here the total kinetic energy KE remains
nearly constant.  A consequence is that with increasing complexity and
increasing CKE along {\sl Path 2}, the DRKE must in turn decrease,
and $\Delta \Omega$ becomes smaller.  This striking
property of achieving a nearly constant KE along {\sl Path 2} (where
both $R_e$ and $P_e$ increase comparably) is a remarkable feature of
this intricate rotating system that is currently unexplained.

Our solutions typically exhibit small differences in behavior in the
two hemispheres, as can be detected in the time-averaged $\Omega$
contours shown in Figure \ref{fig4}, and in the associated latitudinal cuts at
fixed radius displayed in Figure \ref{fig6} for cases $AB$ and $C$.  The
meridional circulations likewise show some symmetry breaking in their
response between the northern and southern hemispheres in Figure \ref{fig7},
which is further quantified for cases $AB$ and $C$ in showing the
meridional streamlines in Figure \ref{fig8} and in examining the latitudinal
variation of the mean velocity component $\hat v_{\theta}$ in Figure
\ref{fig9}.  A sense of these asymmetries can also be assessed by examining
differences in the kinetic energy of differential rotation in the two
hemispheres.  For case $AB$, DRKE in the northern hemisphere is $2.12
\times 10^6$ erg cm$^{-3}$ and $2.09 \times 10^6$ erg cm$^{-3}$ in the
southern hemisphere, or a 1.6\% difference.  For case $C$, the
corresponding values are $1.82 \times 10^6$ and $1.76 \times 10^6$, or
3.6\%.  We expect that such symmetry breaking is likely to evolve
slowly, with neither hemisphere favored.  We plan to study aspects of
symmetry breaking further with more extended simulations in the near
future. Such efforts are inspired by the evolving meridional circulations and mean zonal flows being detected by helioseismology (Haber et al. 2000, 2001), and the differing solar rotation rates in the two hemispheres deduced from
tracking sunspots (Howard, Gilman \& Gilman 1984).

\section{INTERPRETING THE DYNAMICS}

Our shells of rotating compressible convection are very complicated
dynamical systems in terms of the nonlinear feedbacks and couplings
that operate.  It is difficult from first principles to
predict or explain their overall behavior in terms of the differential
rotation and meridional circulations that can be achieved and sustained
as we sample different sites in parameter space.  The five simulations
represent numerical experiments that seek to probe some of the families
of responses within a highly simplified version of the solar convection
zone.  Although most of our approximations here seem reasonable and
necessary to yield a problem tractable to computational experiments, we
do not fully know their impact and thus must draw our interpretations
about the operation of the overall dynamics with considerable caution.
The numerical solutions have the enormous advantage that we can
interrogate them in detail to study various balances and fluxes, and
these help to provide insights about the dynamical system.

\subsection{Redistributing the angular momentum}

Our choice of stress-free boundaries at the top and bottom of the
computational domain has the advantage that no net torque is applied to
our convective shells resulting in the conservation of the angular
momentum. We seek here to identify the main physical processes
responsible for redistributing the angular momentum within our rotating
convective shells, thus yielding the differential rotation seen in
our five cases.  We may assess the transport of angular momentum within
these systems by considering the mean radial (${\cal F}_r$) and
latitudinal (${\cal F}_{\theta}$) angular momentum fluxes.  As
discussed in Elliott et al. (2000), the $\phi$-component of the
momentum equation expressed in conservative form and averaged in time
and longitude yields
\begin{equation}
\frac{1}{r^2} \frac{\p(r^2 {\cal F}_r)}{\p r}+\frac{1}{r \sin\theta}
\frac{\p(\sin \theta {\cal F}_{\theta})}{\p
\theta}=0,
\end{equation}
involving the mean  radial angular momentum flux
\begin{equation}
{\cal F}_r=\hat{\rho}r\sin\theta[-\nu r\frac{\p}{\p
r}\left(\frac{\hat{v}_{\phi}}{r}\right)+\widehat{v_{r}^{'}
v_{\phi}^{'}}+\hat{v}_r(\hat{v}_{\phi}+\Omega_0 r\sin\theta)] \end{equation}
and the mean latitudinal angular momentum flux
\begin{equation}
{\cal F}_{\theta}=\hat{\rho}r\sin\theta[-\nu
\frac{\sin\theta}{r}\frac{\p}{\p
\theta}\left(\frac{\hat{v}_{\phi}}{\sin\theta}\right)+\widehat
{v_{\theta}^{'} v_{\phi}^{'}}+\hat{v}_{\theta}(\hat{v}_{\phi}+\Omega_0
r\sin\theta)].
\end{equation}

In the above expressions for both fluxes, the first terms in each
bracket are related to the  angular  momentum  flux due  to  viscous
transport (which we denote as  ${\cal  F}_{r,V}$ and ${\cal
F}_{\theta,V}$), the second term to the transport due  to  Reynolds
stresses (${\cal F}_{r,R}$ and ${\cal F}_{\theta,R}$) and the  third
term to the transport by the meridional circulation (${\cal F}_{r,M}$
and  ${\cal F}_{\theta,M}$). The Reynolds stresses above are associated
with correlations of the velocity components such as the
$\widehat{v_{r}^{'} v_{\theta}^{'}}$ correlation, which arise from
organized tilts within the convective structures, especially in the
downflow plumes (e.g. Brummell et al. 1998, Miesch et al. 2000).

\begin{figure*}[!ht]
\centerline{\includegraphics[width=1.0\linewidth]{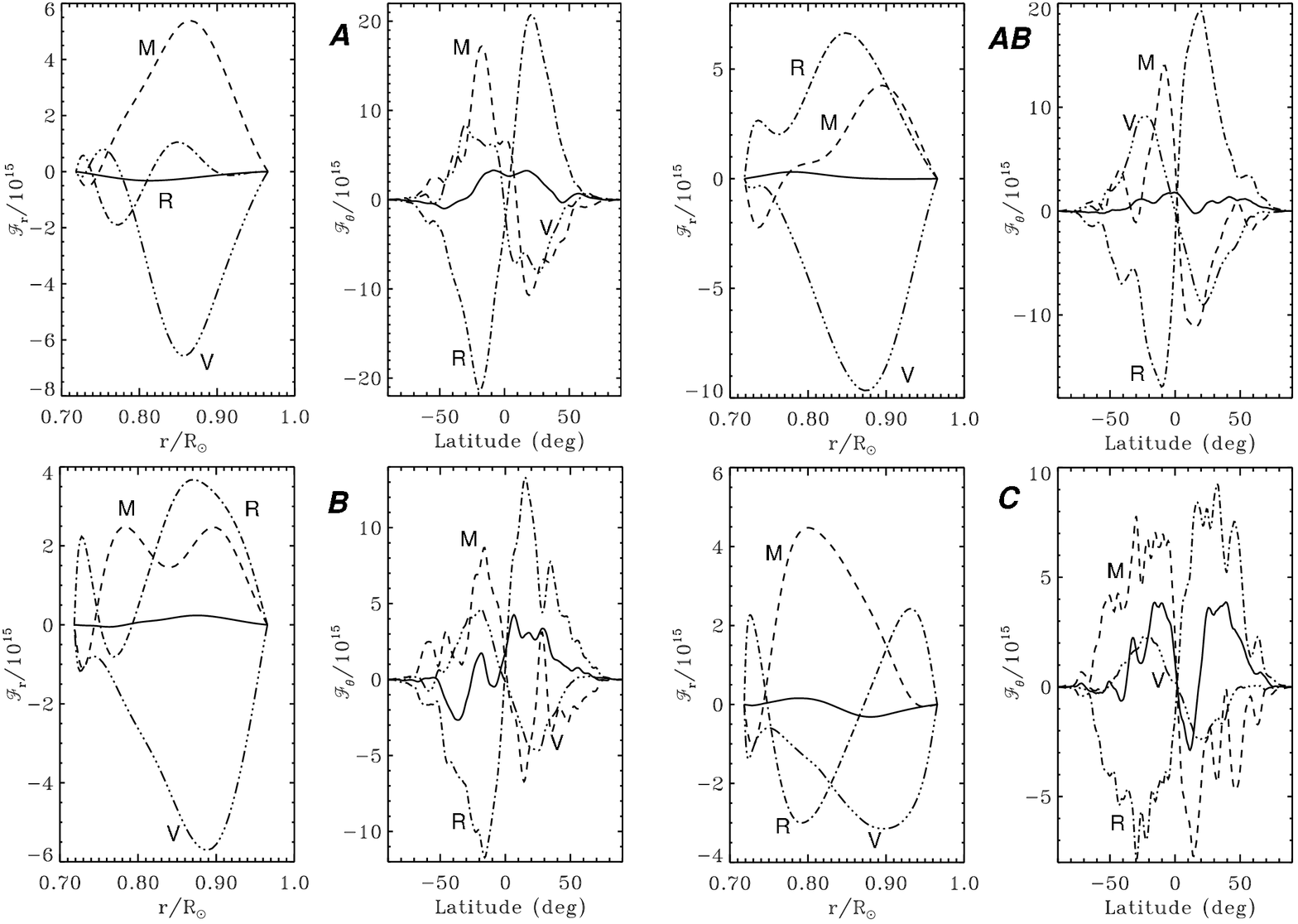}}
\caption[]{\label{fig11} Time average of the latitudinal line integral
of the angular momentum flux ${\cal F}_r$ (left panels in each pair) and
of the radial line
integral of the angular momentum flux ${\cal F}_{\theta}$ (right panels)
for cases $A$ (top left), $AB$ (top right), $B$ (bottom left) and $C$ (bottom
right). The fluxes have been decomposed into their viscous
(labelled V), Reynolds stress (R), and meridional circulation (M) components.
The solid curves represent the total fluxes and serve to indicate the
quality of stationarity achieved.
The positive values represent a radial flux that is directed outward, and a
latitudinal flux directed from north to south. The fluxes for $A$, $AB$, $B$ and $C$
have been averaged over periods in turn of 295, 275, 275 and 175 days.}
\end{figure*}

In Figure \ref{fig11} we show the components of  ${\cal F}_r$ and
${\cal F}_{\theta}$ for cases $A$, $AB$, $B$  and $C$, having
integrated along co-latitude  and radius respectively to deduce the
net  fluxes through shells at various radii and  through cones at
various latitudes, namely in the manner
\begin{equation}
I_{{\cal F}_r}(r)=\int_0^{\pi} {\cal F}_r(r,\theta) \, r^2 \sin\theta
\, d\theta \; \mbox{ , }
\end{equation}
\begin{equation} 
I_{{\cal F}_{\theta}}(\theta)=\int_{r_{bot}}^{r_{top}} {\cal
F}_{\theta}(r,\theta) \, r \sin\theta \, dr \, ,
\end{equation}
and then identify in turn the contributions from viscous (V), Reynolds
stresses (R) and meridional circulation (M) terms. This representation
is helpful in considering the sense and amplitude of the transport of
angular  momentum within the convective shells by each component of
${\cal F}_r$ and ${\cal F}_{\theta}$.

Turning first to the radial fluxes in the leftmost of each pair of
panels in Figure \ref{fig11}, we note that the integrated viscous flux
${\cal F}_{r,V}$ is negative (where for simplicity we drop $I$),
implying a radially inward transport of angular momentum. This property is in
agreement with the positive radial gradient in the angular velocity
profiles achieved in our four cases, as seen in Figure \ref{fig4} in
the radial cuts for different latitudes of $\Omega$. Such downward
transport of angular momentum is well compensated by the two other
terms ${\cal F}_{r,R}$ and  ${\cal F}_{r,M}$, having reached a
statistical equilibrium of nearly no net radial flux, as can be seen by
noting that the solid curve ${\cal F}_r$ is close to zero.  Although
all of our solutions possess complicated temporal variations, our
sampling in time to obtain the averaged fluxes suggest that we are
sensing the equilibrated state reasonably well.  As the level of
turbulence is increased in going from case $A$ to $C$, ${\cal F}_{r,V}$
reduces in amplitude and the transport of angular momentum by the
Reynolds stresses and by the meridional circulation change accordingly
to maintain equilibrium.  The meridional circulation as ${\cal
F}_{r,M}$ involves a strong dominantly outward transport of angular
momentum.  The Reynolds stresses as ${\cal F}_{r,R}$ vacillate in their
sense with depth, though consistently possess outward transport in the
upper portions of the domain.  Case $AB$ is distinguished by ${\cal
F}_{r,R}$ being directed outward throughout the domain.  Detailed
examination with radius and latitude of the Reynolds stress
contributions to the angular momentum fluxes in equations (7--9)
reveals that the `flux streamfunctions' (not shown) possess
multi-celled structures with radius at latitudes above 45$^{\circ}$ for
all cases except $AB$.  This striking difference in case $AB$ of having
a big positive ${\cal F}_{r,R}$, appears to influence the
redistribution of angular momentum at high latitudes.  This may be key
in the monotonic decrease of $\Omega$ with latitude of case $AB$
extending into the polar regions, and provides our first clue
for how {\sl Issue 1} is resolved within this case.  In a broader sense
in considering all of our cases, we deduce that in the radial direction
the transport of angular momentum is significantly affected by both the
meridional circulation and the Reynolds stresses.

The latitudinal transport of angular momentum ${\cal F}_{\theta}$, in
the rightmost of the panels in Figure \ref{fig11}, involves more
complicated and sharper variations in latitude.  This comes about due
to the more intricate latitudinal structure of the differents terms
contributing to the transport.  Here the transport of angular momentum
by Reynolds stresses ${\cal F}_{\theta,R}$ appears to be the dominant
one, being consistently directed toward the equator (i.e. negative in
the south hemisphere and positive in the north hemisphere).  This is an
important feature, since it implies that the equatorial acceleration
observed in our simulations is mainly due to the transport of angular
momentum by the Reynolds stresses, and thus is of dynamical origin.  As
we try to understand {\sl Issue 2}, concerned with retaining a
significant $\Delta \Omega$ as the flow complexity is increased, we
find that the variation of angular momentum fluxes by Reynolds stresses
with increasing complexity along {\sl Paths 1} and {\sl 2} are fairly
similar in character.  Along both these paths the Reynolds stress
fluxes remain prominent, and this appears to sustain the large $\Delta
\Omega$, thereby resolving {\sl Issue 2} for solutions with the level
of turbulence attained in cases $C$ and $D$ (the latter is not shown in
Figure \ref{fig11}, but its transport properties are comparable to those
of case $C$).  Further, we see that the transport by meridional
circulation ${\cal F}_{\theta,M}$ is opposite to ${\cal F}_{\theta,R}$,
with the meridional circulation seeking to slow down the equator and
speed up the poles. A distinguishing feature of case $AB$ is that ${\cal F}_{\theta,M}$
becomes small at latitudes above 30$^{\circ}$, with the tendency of the
meridional circulation to try to spin up the high latitudes sharply
diminished compared to the other cases.  This appears to result from
the strong meridional circulation in case $AB$ being largely confined
to the interval from the equator to 30$^{\circ}$ in latitude (Fig.
\ref{fig8}$a$), with only a weak response at higher latitudes.  This property of
${\cal F}_{\theta,M}$, together with the uniformly positive ${\cal
F}_{r,R}$, provides the second clue for how {\sl Issue 1} appears to be
resolved by case $AB$. As the level of turbulence is increased, we find a
reduction in the amplitudes of all the components of ${\cal
F}_{\theta}$, with ${\cal F}_{\theta,V}$ always being the smallest and
transporting the angular momentum poleward in the same sense as ${\cal
F}_{\theta,M}$.  For ${\cal F}_{\theta,R}$, this lessening amplitude
appears to come about from the increasing complexity of the flows
implying smaller correlations in the Reynolds stress terms, but it is
likely that strengthening coherent turbulent plumes can serve to
rebuild such correlations (Brummell et al. 1998).

Our estimates of the latitudinal transports of angular momentum yield
fairly good equilibration for cases $A$ and $AB$, with little net
latitudinal flux, but the more turbulent cases such as $C$ are
sufficiently complex that achieving such latitudinal balance is a slow
process in the temporal averaging. We conclude that the Reynolds stresses have the dominant role
in achieving the prograde equatorial rotation seen in our simulations,
with its effectiveness limited by the opposing transport of angular
momentum by the meridional circulation. The viscous transports are
becoming more negligible as we achieve more turbulent flows by reducing
the eddy diffusivities.

\begin{figure*}[!ht]
\centerline{\includegraphics[width=0.4\linewidth,angle=90]{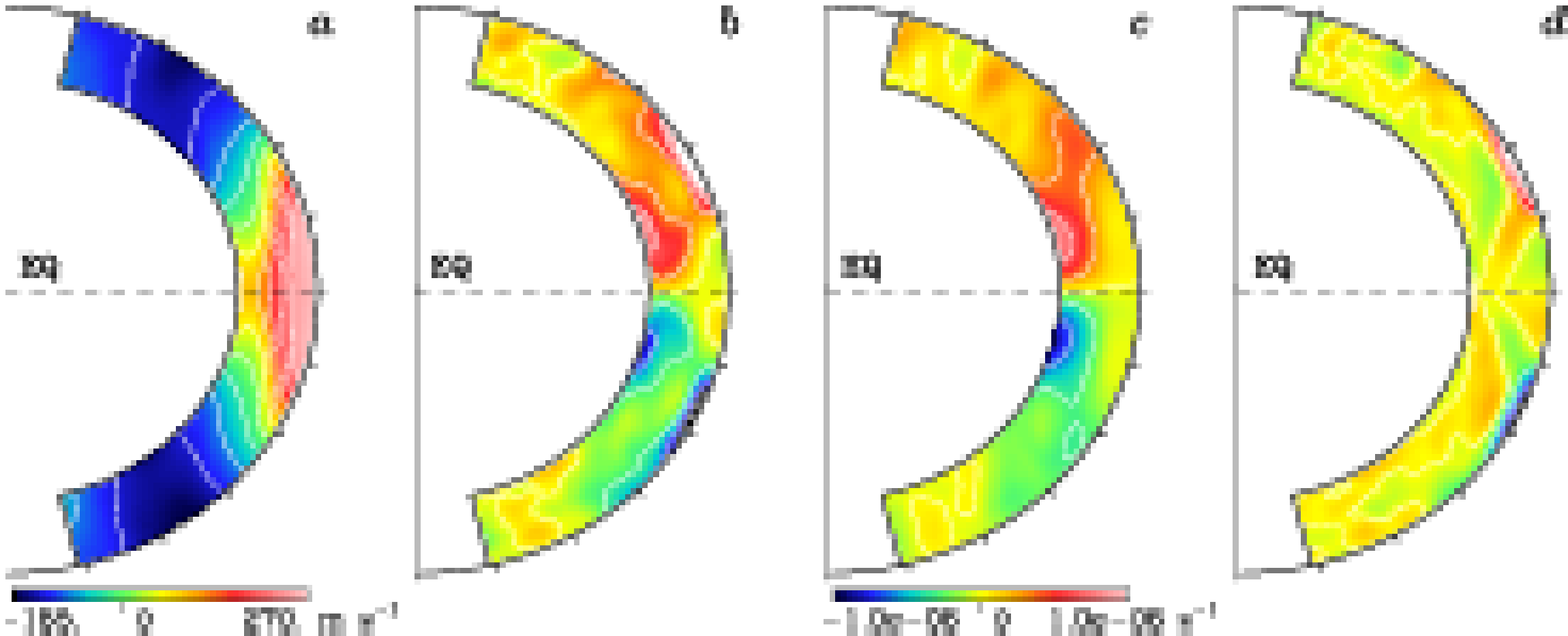}}
\vspace{0.2cm}

\caption[]{\label{fig12} Temporal and longitudinal average for case $AB$ of ($a$)
the longitudinal velocity $\vph$, ($b$) its derivative along the z axis,
$\partial\vph / \partial z$, ($c$) the baroclinic term in the
meridional force balance (see equation 14), and ($d$) the difference
between the last two terms [namely ($b$) minus ($c$)].  The results
have been averaged over a period of 275 days. Panel ($d$) shows
that there are major departures  from a simple thermal-wind balance,  
especially near the top of the domain. The same color scale is used in panels
($b$), ($c$) and ($d$).}
\end{figure*}

\subsection{Baroclinicity and thermal winds}

Convection influenced by rotation can lead to latitudinal heat
transport in addition to radial transport, thereby producing
latitudinal gradients in temperature and entropy even if none were
imposed by the boundary conditions.  This further implies that surfaces
of constant mean density and mean pressure will not coincide, thereby
admitting baroclinic terms in the vorticity equations (Pedlosky 1987,
Zahn 1992).  Baroclinicity has been argued to possibly have a pivotal
role in obtaining differential rotation profiles whose angular
velocity, like the sun, are not constant on cylinders (e.g.
Kitchatinov \& R\"udiger 1995).  We shall here analyze our cases $AB$
and $C$ from that perspective, finding that though a small latitudinal
entropy gradient is realized, the resulting differential rotation as
exhibited in our solutions by the mean longitudinal velocity
$\hat{v}_{\phi}$ cannot be accounted for principally by the baroclinic
term.   To make such interpretation specific, we should turn as in
Elliott et al. (2000) to the mean (averaged in longitude and time)
zonal component of the curl of the momentum equation (2), expressed as
\begin{eqnarray} \label{rf}
\epsilon_{\phi ij}
\frac{\partial}{\partial x_{i}}\left(\widehat{v_{k}\frac{\partial
v_{j}}{\partial x_{k}}}+\hat{\rho}\,^{-1}\frac{\partial}{\partial
x_{k}}\widehat{{\cal D}_{kj}}\right) &=& 2\,\Omega_o\,\frac{\partial
\hat{v}_{\phi}}{\partial z} \\
&+&\hat{\rho}\,^{-2}\del\,\hat{\rho}\,\dotp\del\,\hat{p}
\Bigr|_{\phi}\mbox{ ,} \nonumber
\end{eqnarray}
where the Einstein summation convention has been adopted, $\epsilon$
represents the permutation tensor, and
\[\frac{\partial}{\partial z}
\equiv \cos\theta \frac{\partial}{\partial r} - \frac{\sin\theta}{r}
\frac{\partial}{\partial \theta} \]
is the derivative parallel to the rotation axis.  This vorticity
equation is helpful in examining the relative importance of different
forces in meridional planes; here terms arising from Reynolds and
viscous stresses are on the left and from Coriolis and baroclinic
effects on the right.  If one were to simply neglect the Reynolds and
viscous stresses, we obtain the simplest version of a `thermal--wind
balance' in which departures of zonal winds from being constant on
cylinders aligned with the rotation axis are accounted for by the
baroclinic term involving crossed gradients of density and pressure,
namely
\begin{equation} 2\,\Omega_o\,\frac{\partial
\hat{v}_{\phi}}{\partial z} = -
\hat{\rho}\,^{-2}\del\,\hat{\rho}\,{\vec{\times}}\del\,\hat{p}\Bigr|_{\phi}\mbox{
.}\label{tw1}\end{equation}
With the superadiabatic gradient expressed as
\begin{equation} \frac{1}{c_{P}}\del\,\hat{S} = \frac{1}{\gamma\hat{p}}
\del\,\hat{p} - \frac{1}{\hat{\rho}}\del\,\hat{\rho}\mbox{
,}\end{equation}
where $\gamma$ is the logarithmic derivative of pressure with respect
to density at constant specific entropy, we can rewrite equation (\ref{tw1})
as
\begin{equation}
\frac{\partial\hat{v}_{\phi}}{\partial z} =
\frac{1}{2
\Omega_o\hat{\rho}c_{P}}\del\,\hat{S}\,\dotp\del\,\hat{p}\Bigr|_{\phi}
= \frac{g}{2 \Omega_o r\,c_{P}} \frac{\partial\hat{S}}{\partial \theta}
\mbox{ ,}\label{tw2}\end{equation}
having neglected turbulent pressure.  Thus breaking the Taylor-Proudman
constraint that requires rotation to be constant on cylinders, with
$\partial\hat{v}_{\phi}/\partial z$ zero, can be achieved by
establishing a latitudinal entropy gradient.  However, Reynolds and
viscous stresses can also serve to break that constraint, and indeed we
next show that those terms are at least as important as the baroclinic
term.

\begin{figure*}[!ht] \setlength{\unitlength}{1.0cm}
\centerline{\includegraphics[width=0.4\linewidth,angle=90]{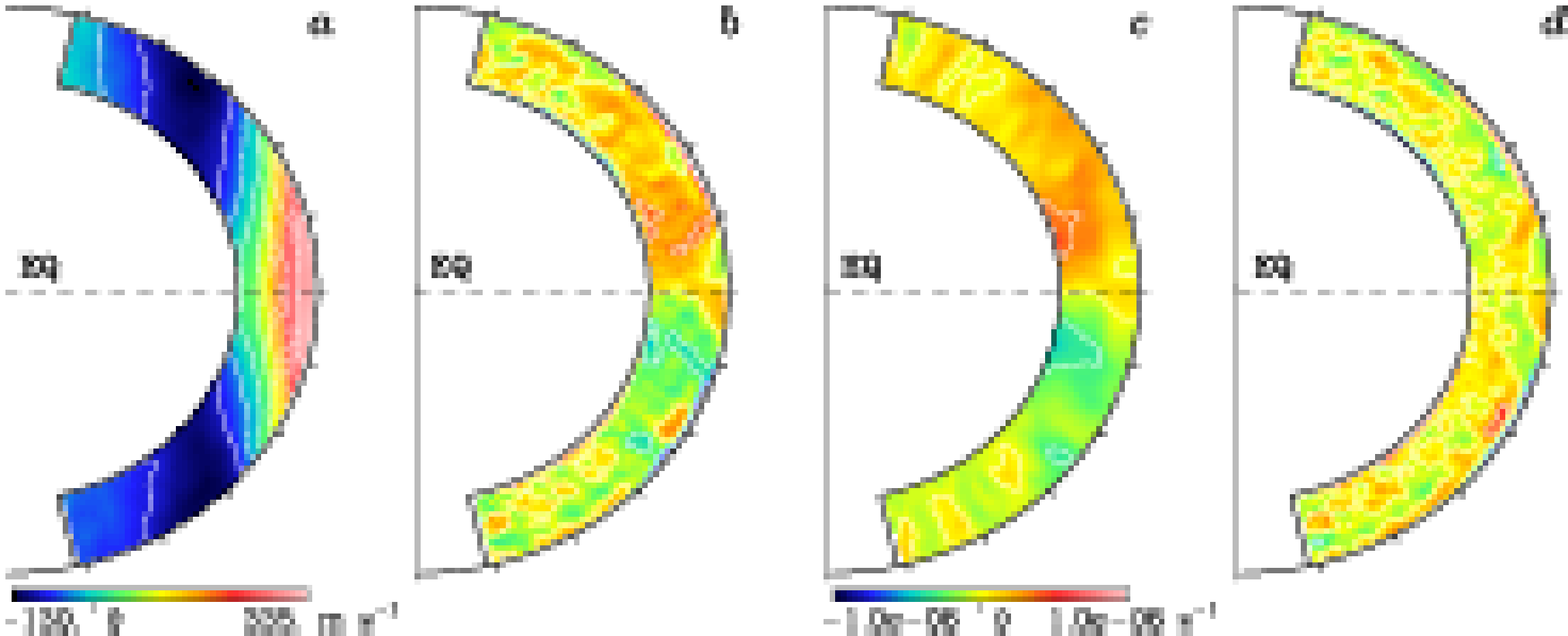}}
\vspace{0.2cm}

\caption[]{\label{fig13} As in Fig. \ref{fig12}, analyzing the role of baroclinicity 
in the more turbulent case $C$ in maintaining the differential rotation.
There are significant departures from a thermal-wind balance in thin regions near the
top and bottom of the shell.}
\end{figure*}

We turn in Figure \ref{fig12} to an analysis of case $AB$ in terms of
how well is a simple thermal--wind balance achieved or violated.
Figures \ref{fig12}$a,b$ display the temporal mean zonal velocity
$\hat{v}_{\phi}$ and its gradient $\partial\hat{v}_{\phi} / \partial
z$, with the latter having pronounced variations at mid latitudes near
the top of the spherical shell and others at lower latitudes near the
bottom of the domain.  The baroclinic term [as on right of equation
({\ref{tw2})] is shown in Figure \ref{fig12}$c$, possessing the largest
amplitudes close to the base of the shell at low latitudes, with a
tongue connecting to mid latitudes in traversing the shell.  The
difference between this baroclinic term and the actual
$\partial\hat{v}_{\phi}/\partial z$, as shown in Figure \ref{fig12}$d$,
is a measure of the effectiveness of a thermal--wind balance in case
$AB$.  It is evident that baroclinicity yields a fair semblance of a
balance over much of the deeper layer, with the baroclinic term (Fig.
\ref{fig12}$c$) typically being greater in amplitude than
$\partial\hat{v}_{\phi}/\partial z$ (Fig. \ref{fig12}$b$) there.
However, the major regions of departure with opposite signs in the two
hemispheres show that in the upper domain, between latitudes of about
15$^\circ$ and 45$^\circ$, that balance is quite severely violated:
there the Reynolds stress terms in equation (\ref{rf}) involving vortex
tube stretching and tilting become the main players.  This broad site
coincides with regions of strong latitudinal gradient in $\hat
v_{\phi}$, and is centered in latitude where the relative rotation
changes sense from prograde to retrograde.  What we have learned from
this is that whereas the convection does establish a latitudinal
gradient of entropy that is needed for baroclinic terms to achieve
aspects of thermal--wind balance over the deeper portions of the
domain, the Reynolds stresses have an equally crucial role in the
meridional force balance over portions of the upper domain. The more
turbulent case $C$ is likewise analyzed in Figure \ref{fig13}, and it
generally exhibits comparable behavior.  The baroclinic term (Fig.
\ref{fig13}$c$) captures much of the $\partial\hat{v}_{\phi}/\partial
z$ variation (Fig. \ref{fig13}$b$) at mid latitudes over most of the
deep shell, but there are large departures (Fig. \ref{fig13}$d$) in thin
domains near the top and bottom of the shell, again between 15$^\circ$
and 45$^\circ$ in latitude.  Thus here too the Reynolds stress terms
are significant players in the overall balance.

\begin{figure*}[!ht]
\centerline{\includegraphics[width=0.95\linewidth]{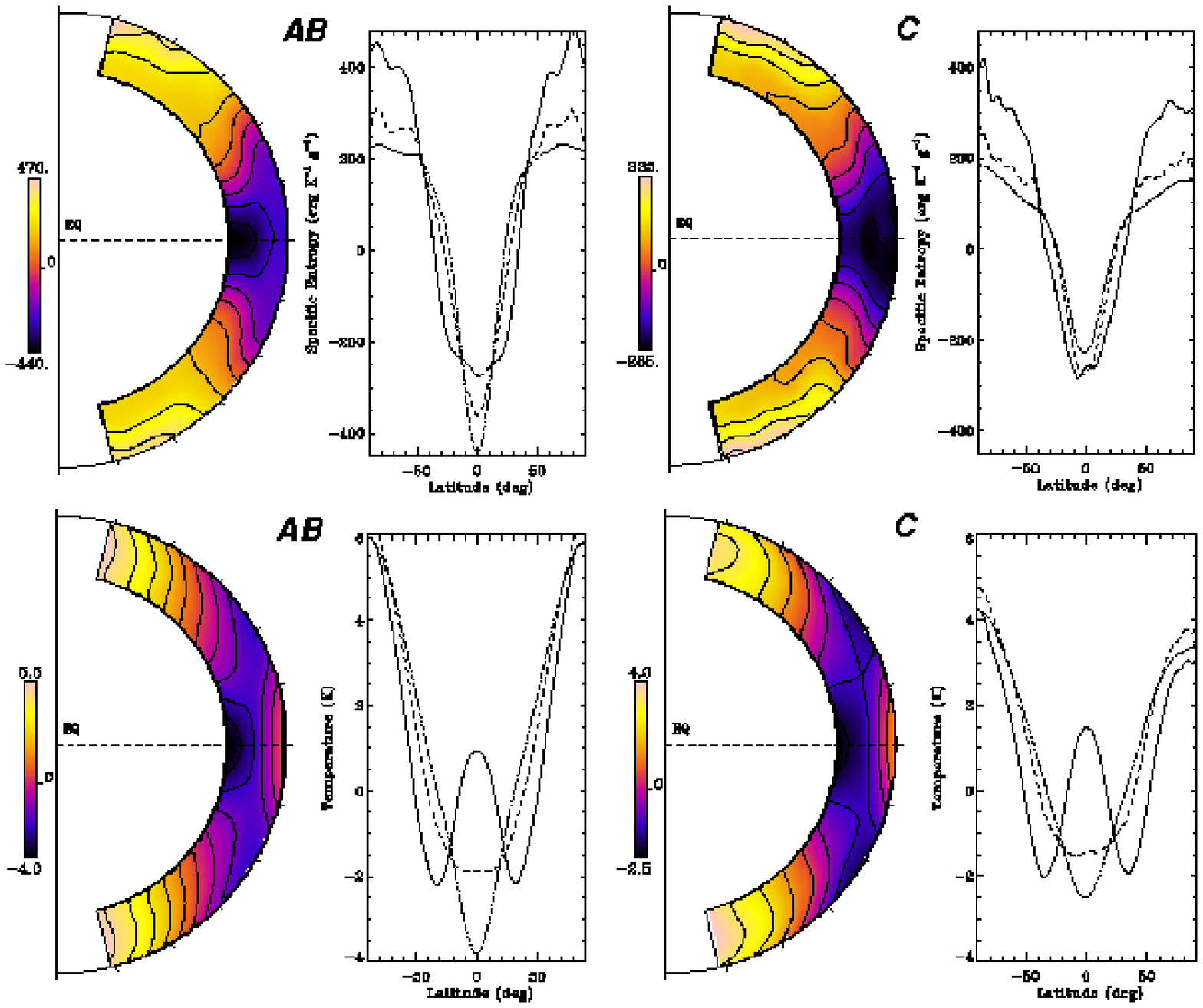}}

\caption[]{\label{fig14} Temporal and longitudinal averages for cases $AB$ and $C$ of the
specific entropy (upper panels) and
temperature fluctuations (lower panels), accompanied by latitudinal profiles at the base
({\it dash dotted} line), at the middle ({\it dashed}) and at the top ({\it solid}) of the
convective domain. The results have been averaged over periods
in turn of 275 and 175 days.  The presence of a latitudinal
variation of entropy is consistent with the baroclinic term (shown in
Figs. \ref{fig12} and \ref{fig13}) and involves an equator to
pole temperature contrast of at most 4 to 8 K near the top where the mean 
temperature is about $10^5$ K.}
\end{figure*}

The latitudinal entropy and temperature gradients established within
our simulations should be examined further.  We show in Figure
\ref{fig14} the time and longitude averaged specific entropy fluctuations
$\hat S$ and temperature fluctuations $\hat T$ for cases $AB$ and $C$,
presenting both color contour renderings across the shell and their
variations with latitude at three depths.  Our model $AB$, which
exhibits the strongest differential rotation, also possesses the
greatest temperature and entropy contrasts with latitude.  We see from
the latitudinal cuts of temperature that a $\Delta\Omega$ of order 30\%
involves a pole--equator temperature variation of about 4 to 8 degree
K, the pole being warmer. These temperature contrasts are very small compared
to the mean temperature near the top of our domain of about $10^5$ K, and
of $2\times 10^6$ K near its base. There is some evidence of a latitudinal
variation in the photospheric temperature of at least a few K with the same 
sense obtained from observations of
the solar limb (e.g. Kuhn 1998), though relative variations of such
small amplitude are very difficult to measure.  We note that our
temperature fields show some banding with latitude near the top of the
domain, with the equator slightly warm, then attaining relatively cool
values with minima at about latitude 35$^\circ$, followed by rapid
ascent to warm values at high latitudes.  The behavior is monotonic
with latitude at greater depths, as it is consistently so for entropy
at all depths.  These differences between temperature and entropy are
accounted for by effects of the pressure field necessary to drive the
meridional circulation.

In summary, although our solutions attain close to a thermal--wind
balance over large portions of the domain, the departures elsewhere are
most significant.  These arise from the Reynolds stresses that have a
crucial role in establishing the differential rotation profiles
realized in our simulations.  The baroclinicity in our solutions,
resulting from latitudinal heat transport that sets up a
pole-to-equator temperature and entropy contrast, contributes to
$\Omega$ not being constant on cylinders, but it is not the dominant
player as envisioned in some discussions of mean-field models of solar
differential rotation (e.g. Kitchatinov \& R\"udiger 1995, Rekowski \&
R\"udiger 1998; Durney 1999, 2000).

\section{CONCLUSIONS}

Our five simulations studying the coupling of turbulent convection and
rotation within full spherical shells have revealed that strong
differential rotation contrasts can be achieved for a range of
parameter values.  With these new models, we have focused on two
fundamental issues raised in comparing the solar differential rotation
deduced from helioseismology with the profiles achieved in the prior
3--D simulations of turbulent convection with the ASH code (Miesch et
al. 2000, Elliott et al. 2000).  As {\sl Issue 1}, the sun appears to
possess remarkably slow poles, with $\Omega$ decreasing steadily with
latitude even at mid and high latitudes (Fig. \ref{fig1}).  In contrast, the
previous models showed little variation in $\Omega$ at the higher
latitudes, having achieved most of their latitudinal angular velocity contrast
$\Delta \Omega$ in going from the equator to about 45$^\circ$.  As {\sl
Issue 2}, there was a tendency for $\Delta \Omega$ to diminish or even
decrease sharply within the prior simulations as the convection became
more turbulent, yielding values of $\Delta \Omega$ that were becoming
small compared to the helioseismic deductions.  In seeking to resolve
these two issues, we have explored two paths in parameter space that
yield complex and turbulent states of convection.  {\sl Path 1}
involves decreasing the Prandtl number in the sequence of cases $A$,
$B$ and $C$, while keeping the P\'eclet number nearly constant.  {\sl
Path 2} maintains a constant Prandtl number as both the Reynolds and
P\'eclet number are increased in the sequence of cases $AB$, $B$ and
$D$.  On both paths the convective Rossby number has been chosen to be
less than unity, thereby maintaining a strong rotational influence on
the convection even as the flows become more intricate.

In dealing with {\sl Issue 1}, our case $AB$ provides the first
indications that it is possible to attain solutions in which the polar
regions rotate significantly slower than the mid latitudes (Fig. \ref{fig4}).
There is a monotonic decrease from the fast (prograde) equatorial rate
in $\Omega$ to the slow (retrograde) rate of the polar regions.
Further, that case $AB$ has $\Omega$ nearly constant on radial lines at
the higher latitudes, again in the spirit of the helioseismic
inferences.  We do not fully understand why in case $AB$ such a
strikingly different $\Omega$ profile results, compared to that in our
other solutions (and of the progenitor simulations) in which the
contrast $\Delta \Omega$ is mainly achieved in the lower latitudes.
Our principal clues come from Figure \ref{fig11} where we find that only in case
$AB$ is the Reynolds stress component of the net radial angular
momentum flux ${\cal F}_{r,R}$ (through shells at various radii)
uniformly directed outward.  From having examined in detail angular
momentum flux streamfunctions (not shown) with radius and latitude
consistent with equations (7--9), we observed that the Reynolds stress
contributions to such transport possessed multi-celled structures with
radius at high latitudes in all the cases except $AB$.  The single-cell
behavior there for case $AB$ appears to enable more effective
extraction of angular momentum by Reynolds stresses from the high to
the low latitudes, thereby yielding a distinctive rotational slowing of
the high latitudes. Further, case $AB$ possesses strong meridional circulations at low
latitudes, but only feeble ones at latitudes above 30$^{\circ}$, unlike
other solutions such as case $C$ (Figs. \ref{fig8}, \ref{fig9}).  This yields a weak
meridional component ${\cal F}_{\theta,M}$ (seeking to spin up the
poles) to the latitudinal angular momentum flux, thereby allowing the
equatorward transport by the Reynolds stress component ${\cal
F}_{\theta,R}$ to succeed in extracting angular momentum from the
higher latitudes. Such polar slowing also leads to case $AB$
possessing the greatest $\Delta \Omega$ attained in our five
simulations (Table $2$).

We also considered the possibility that the slow pole behavior in case
$AB$ may have baroclinic origins.  This can result from suitable
correlations in velocity and thermal structures yielding a latitudinal
heat flux which may produce substantial entropy variations at the
higher latitudes, thereby leading to greater baroclinic contributions
in equation (11) that defines the variation of mean zonal velocity.
Examination of Figure \ref{fig12} at high latitudes does not reveal a
prominent baroclinic contribution, and this is consistent with the
bland variation of entropy for case $AB$ (Fig. \ref{fig14}) at latitudes
above about 40$^\circ$.  We conclude that the origin of the slow rotation
rate in polar regions appears to be primarily dynamical, being
associated with the Reynolds stress transports, and not with
baroclinicity that arises from latitudinal heat transport
serving to establish a sufficiently warm pole.  Although case $AB$
provides a solution that resolves {\sl Issue 1}, it is unique in
achieving this among our five simulations.  It may be that in parameter
space there only exists a small basin of attraction for such behavior,
though we think it more likely that several solution states may coexist
for the same control parameters, one of which exhibits the gradual
rotational slowing at high latitudes, and others having most $\Omega$
variations confined to low and mid latitudes.  We plan to examine
whether the slow pole characteristics of case $AB$ can be maintained at
nearby sites in parameter space if started from initial conditions
corresponding to $AB$, and plan to report on this in the future.

{\sl Issue 2} concerns sustaining a strong differential rotation with
latitude as the convection becomes more complex.  The two paths that we
have explored in parameter space to achieve more complex and turbulent
states yield relative angular velocity contrasts $\Delta \Omega/ \Omega_o$ in
latitude that are comparable to values deduced from helioseismology,
with both our models and apparently the sun possessing a contrast of
order 30\%.  Further, this is accomplished while imposing an upper
thermal boundary condition that ensures a uniform emerging heat flux
with latitude, as suggested in Elliott et al. (2000).  {\sl Path 1}
involving a decreasing Prandtl number is somewhat more effective in
attaining large $\Delta \Omega$ as the solutions become turbulent than
{\sl Path 2} which has the Prandtl number fixed at 0.25 as both
diffusivities are decreased.  This holds out hope that even more
turbulent solutions will act likewise.

We have shown that the strong $\Delta \Omega$ results from the role of
the Reynolds stresses in redistributing the angular momentum.  This
transport is established by correlations in velocity components arising
from convective structures that are tilted toward the rotation axis and
depart from the local radial direction and away from the meridional
plane.  These yield both $v_r v_{\phi}$ and $v_{\theta} v _{\phi}$
correlations necessary for the Reynolds stress contributions to the
radial and latitudinal angular momentum fluxes analyzed in Figure \ref{fig11}.
The fast downflow plumes have a dominant role in such Reynolds
stresses, much as seen in local studies (Brummell et al. 1998).  Our
simulations have attained a spatial resolution adequate to begin to
attain coherent structures amidst the turbulence, which is believed to
be a key in sustaining strong Reynolds stresses at higher turbulence
levels.  This has the consequence that all our spherical shells possess
fast prograde equatorial rotation relative to the reference frame.
There are some contributions toward maintaining the differential
rotation from the latitudinal heat transport inherent in our convection
that serves to establish a warm pole (with a contrast of a few K)
relative to the equator, with baroclinicity and a partial thermal--wind
balance helping to yield equatorial acceleration.  The meridional
circulations generally work to oppose such tendencies by redistributing
angular momentum so as to try to spin up the poles.  Our simulations on
{\sl Paths 1} and {\sl 2} confirm that strong differential rotation
with fast equators has its primary origin in angular momentum transport
associated with the Reynolds stresses.  Such prominent transports serve
to resolve {\sl Issue 2}.  Our next challenge is to satisfy {\sl Issue 1}
simultaneously with {\sl Issue 2} in the more turbulent solutions, which may
also lead to $\Omega$ being more nearly constant on radial lines at mid to
high latitudes.

Although our results for $\Omega$ have made promising contacts with
helioseismic deductions about the state of solar differential rotation
in the bulk of the convection zone, there are also major issues that we
have not yet tackled.  We must evaluate more advanced subgrid-scale
terms in representing the unresolved turbulence within such
simulations, especially in the near-surface regions.  Foremost are also
questions of how does the presence of a region of penetration below the
convection zone influence the angular momentum redistribution in the
primary zone above, and does the tachocline of shear that is
established near the interface with the deeper radiative interior
modify properties within the convection zone itself.  We are keen to
also investigate aspects of the rotational shear evident close to the
solar surface.  This is just now becoming computationally feasible, and
involves extending our computational domain upward and beginning to
resolve supergranular motions there, as discussed in DeRosa \& Toomre
(2001) in preliminary studies with thin shells.  We are still at early
stages with our simulations using ASH to study turbulent convection in
spherical shells, yet it is comforting that the mean differential
rotation profiles realized in some of the simulations are beginning to
capture many of the dominant features for $\Omega$ deduced from the
helioseismic probing.

We thank Nicholas Brummell, Marc DeRosa, Julian Elliott, Peter Gilman,
Mark Miesch and Jean-Paul Zahn for useful discussions and comments
during the writing phase of this paper, and a referee for encouraging
us to clarify the objectives and thrust of the presentation.  This work
was partly supported by NASA through SEC Theory Program grant NAG5-8133
and by NSF through grant ATM-9731676.  Various phases of the
simulations with ASH were carried out with NSF PACI support of the San
Diego Supercomputer Center (SDSC), the National Center for
Supercomputing Applications (NCSA), and the Pittsburgh Supercomputing
Center (PSC).  Much of the analysis of the extensive data sets was
carried out in the Laboratory for Computational Dynamics (LCD) within
JILA.}

\pagebreak

\begin{table*}[!ht]
\begin{center}
\caption[]{Parameters for the Five Simulations}
\vspace{0.2cm}
\begin{tabular}{||c||ccccc||}
\tableline
\tableline
 Case & {\sl A} & {\sl AB} & {\sl B} & {\sl C} & {\sl D} \\
\tableline
\tableline
  $N_r, N_{\theta}, N_{\phi}$ & 64, 128, 256 & 64, 128, 256 & 64, 256, 512 &
192, 256, 512  & 192, 512, 1024\\
  $R_a$ & 3.1 $\times 10^4$ & 3.4 $\times 10^4$ & 1.4 $\times 10^5$ & 3.1 $\times 10^5$& 6.5
$\times 10^5$\\
  $T_a$  & 7.7 $\times 10^4$ & 3.1 $\times 10^5$ & 1.2 $\times 10^6$ & 5.4 $\times 10^6$& 6.5
$\times 10^6$\\
  $P_r$ & 1 & 0.25 & 0.25 & 0.125 & 0.25\\
  $R_{oc}$ & 0.645 & 0.662 & 0.673 &0.682  & 0.633\\
  $\nu$  & 5.5 $\times 10^{12}$ & 2.8 $\times 10^{12}$&1.4 $\times 10^{12}$ & 6.8 $\times 10^{11}$ &
6.0 $\times 10^{11}$\\
  $\kappa$  &5.5 $\times 10^{12}$ & 1.1 $\times 10^{13}$ & 5.5 $\times 10^{12}$ & 5.5 $\times 10^{12}$
& 2.4 $\times 10^{12}$\\
  $\tilde{R}_e$  & 28 & 85 & 
170&385 &410\\
  $\tilde{R}_o$  & 0.10 & 0.16 & 0.15
& 0.17 & 0.16\\
  $\tilde{P}_e$ & 28 & 21 & 43
& 48 & 103\\
 \tableline
 \tableline
\end{tabular}
\end{center}

All simulations have an inner radius $r_{bot}=5.0 \times 10^{10}$ cm, 
an outer radius $r_{top}=6.72 \times 10^{10}$ cm, with 
$L=1.72 \times 10^{10}$ cm the thickness of the computational domain. 
The number of radial, latitudinal and longitudinal mesh points are $N_{r}, N_{\theta}, N_{\phi}$.
Here evaluated at mid-layer depth are the Rayleigh number $R_a=(-\p
\rho/\p S)\Delta S g L^3/\rho \nu \kappa$, the Taylor number $T_a=4
\Omega^2 L^4/\nu^2$, the Prandtl number $P_r=\nu/\kappa$, the
convective Rossby number $R_{oc}=\sqrt{R_a/T_a P_r}$, the rms Reynolds number
$\tilde{R}_e=\vvr L/\nu$, the rms P\'eclet number
$\tilde{P}_e=\tilde{R}_eP_r=\vvr L/\kappa$, and the rms Rossby
number $\tilde{R}_o=\tilde{\omega}/2\Omega\sim\tilde{v}/2\Omega L$,
where $\vvr$ is a representative rms convective velocity. A Reynolds number based on the
peak velocity at mid depth would be about a factor 4 larger. The eddy viscosity $\nu$
and eddy conductivity $\kappa$ at mid depth are quoted in cm$^2\,$s$^{-1}$.

\end{table*}

\begin{table*}[!ht]
{\footnotesize
\begin{center}
\caption[]{Representative Velocities, Energies and Differential Rotation}\label{Table 3b}
\begin{tabular*}{1.1\linewidth}{@{\extracolsep\fill}||p{0.8cm}||p{0.4cm}p{0.4cm}p{0.4cm}p{0.4cm}p{0.4cm}p{0.4cm}|p{1.cm}p{2.1cm}p{2.1cm}p{2.4cm}|p{1.0cm}||}
\tableline
\tableline
\multicolumn{1}{||c||}{}&\multicolumn{6}{c|}{Mid Convective Zone}&
\multicolumn{4}{c|}{Volume Average} &\\
\multicolumn{1}{||c||}{Case}& $\vrr$ & $\vtr$ & $\vphr$ & $\vphr'$ & $\vvr$ &
$\vvr'$ &
\multicolumn{1}{c}{KE} & DRKE & CKE & MCKE & $\Delta\Omega/\Omega_o$\\
\tableline
\tableline
\multicolumn{1}{||c||}{\sl A} & 46 & 40& 69& 44& 92&74 & 2.7$\times 10^6$& 8.2$\times 10^5$ (30\%) & 1.9$\times 10^6$ (70\%)&$1.0\times 10^4$ (0.37\%) & 12\% \\
\multicolumn{1}{||c||}{\sl AB} & 50 & 47& 124& 53& 142&87 & 6.5$\times 10^6$& 4.2$\times 10^6$ (64\%)& 2.3$\times 10^6$ (36\%) &2.1$\times 10^4$ (0.32\%) & 33\% \\
\multicolumn{1}{||c||}{\sl B} & 57&56 &115 &59 &140 &99 & 6.5$\times 10^6$ &3.4$\times 10^6$ (52\%)& 3.1$\times 10^6$ (48\%)& 2.5$\times 10^4$ (0.38\%) & 28\% \\
\multicolumn{1}{||c||}{\sl C} &68 &67 &122 &70 &155 &117 &7.9$\times 10^6$ & 3.6$\times 10^6$ (46\%)& 4.3$\times 10^6$ (54\%) &3.3$\times 10^4$ (0.42\%) & 30\% \\
\multicolumn{1}{||c||}{\sl D} &72 &67 & 108& 64&146 &111 &6.5$\times 10^6$ & 2.3$\times 10^6$ (35\%)& 4.2$\times 10^6$ (65\%) & 3.0$\times 10^4$ (0.46\%) & 25\% \\
 \tableline
 \tableline
\end{tabular*}
\end{center}}

In the five cases, temporal averages at mid-layer depth in convection zone of rms components of
velocity $\vrr$, $\vtr$, $\vphr$ and of speed $\vvr$, and of fluctuating velocities $\vphr'$ and $\vvr'$  (with temporal and azimuthal mean subtracted), all expressed in m$\,$s$^{-1}$. Also
listed are the time averages over the full domain of the total kinetic energy, KE, that
associated with the (axisymmetric) differential rotation, DRKE, with the (axisymmetric) 
meridional circulation, MRKE, and with the non-axisymmetric convection itself, CKE, all in
erg$\,$cm$^{-3}$. The relative latitudinal contrast of angular velocity $\Delta\Omega/\Omega_o$
between $0^{\circ}$ and $60^{\circ}$ near the top of the domain are stated.

\end{table*}

\end{document}